\documentclass[prb,amsmath,superscriptaddress,amssymb,twocolumn,showpacs]{revtex4}
\usepackage{graphicx}
\usepackage{bm}
\usepackage{color}
\usepackage{dsfont}
\DeclareMathOperator{\Tr}{Tr}
\DeclareMathOperator{\Pf}{Pf}
\begin{document}

\title{Excess charges as a probe of one-dimensional topological crystalline insulating phases}
\author{Guido van Miert}
\affiliation{Institute for Theoretical Physics, Centre for Extreme Matter and Emergent Phenomena, Utrecht University, Princetonplein 5, 3584 CC Utrecht, The Netherlands}
\author{Carmine Ortix}
\affiliation{Institute for Theoretical Physics, Centre for Extreme Matter and Emergent Phenomena, Utrecht University, Princetonplein 5, 3584 CC Utrecht, The Netherlands}
\affiliation{Institute for Theoretical Solid State Physics, IFW Dresden, PF 270116, 01171 Dresden, Germany}

\begin{abstract}
We show that in conventional one-dimensional insulators excess charges created close to the boundaries of the system can be expressed in terms of the Berry phases associated with the electronic Bloch wave functions. Using this correspondence, we uncover a link between excess charges and the topological invariants of the recently classified one-dimensional topological phases protected by spatial symmetries. Excess charges can be thus used as a probe of crystalline topologies.
\end{abstract} 

\maketitle
\section{Introduction}
\label{sec:secintro}
The electronic properties of both insulating and metallic crystals are largely characterized by the electronic band structure: It relates the crystal momentum $q$ to the corresponding energy $E_n(q)$, with $n$ the band index. In an insulator the Fermi level lies in a band gap separating the conduction from the valence bands, whereas in metals the Fermi level intersects an energy-momentum curve. Although the bulk band structure plays an indispensable role in describing optical, magnetic, and electrical properties of materials, it does not describe all the relevant electronic properties, even in a simple single-particle picture. Edge effects are a notable example. 

In topological states of matter the presence of metallic edge states mandated by topology cannot be extracted from the bulk band structure.\cite{HasanKane,QiZhang} Surface Dirac cones in three-dimensional (3D) topological insulators\cite{FuKaneMele,JoelMoore,Fukane2,SCZHANG1,SCZHANG2,VDBRINK} and crystalline topological insulators,\cite{LFU1,LFU2,Tanaka,VJ,AndoFu} Fermi arcs in 3D Weyl semimetals,\cite{BalentsBurkov,Savrasov,Hasan,Ding,Hasan2,Lau2} chiral and helical edge states in two-dimensional (2D) insulators\cite{vonklitzing,KaneMele,KaneMele2,BHZ,Molenkamp} 
are all exceptional features escaping the conventional bulk band structure picture. These are instead encoded in the bulk Hamiltonian.
Being a completely general phenomenon, however, edge effects inevitably appear also in metals as well as in insulating states of matter which are topologically trivial according to the Altland-Zirnbauer classification.\cite{AZ,Schnyder,Kitaev,Ryu}In particular, this applies to one-dimensional (1D) insulators that do not carry a topological invariant in the absence of particle-hole and chiral symmetry.  In these systems, the presence of edges does not yield additional localized metallic modes but the different boundary conditions do affect the wave function.

\begin{figure}[tbp]
\centering
\includegraphics[width=.46\textwidth]{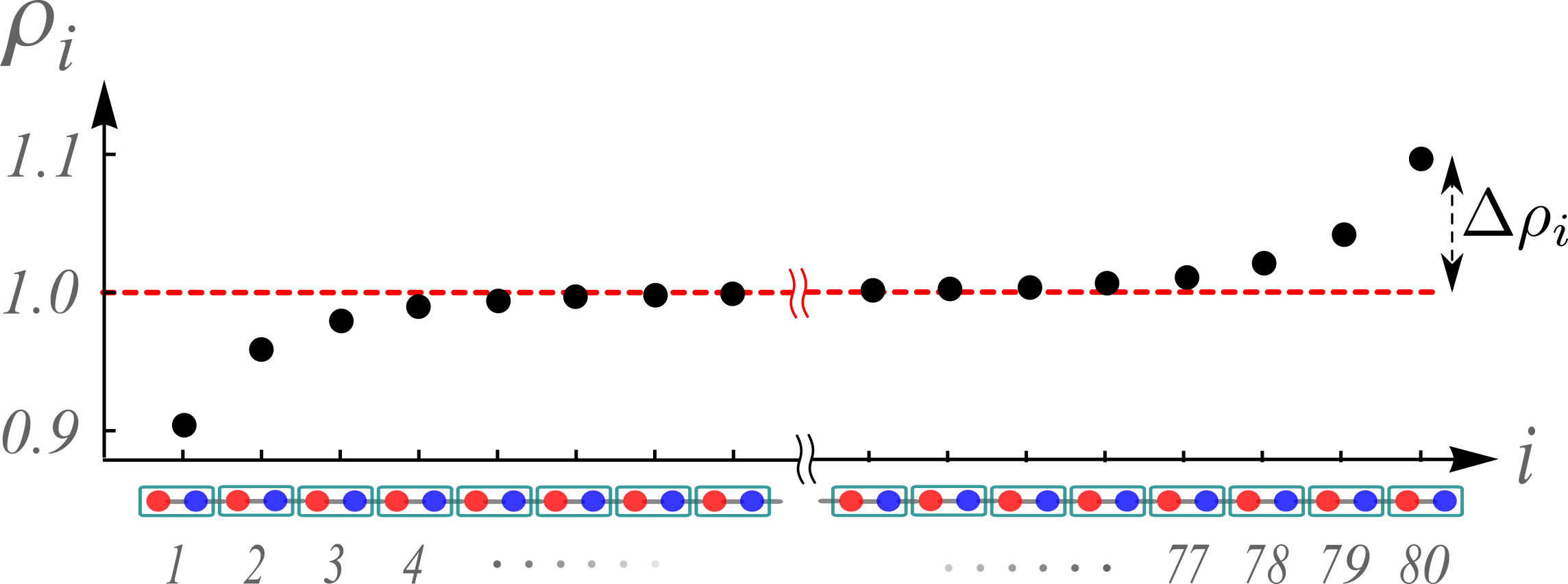}
\caption{Charge density per unit-cell $\rho_i$ for a binary chain with open boundary conditions and $M=80$: the green rectangle denotes the preferred unit-cell, the red (blue) sites have onsite energies $m$($-m$), and the hopping parameter is given by $t$, which we assume to be positive. Here we used $m/t=0.25$.}
\label{fig:fig1}
\end{figure}

Put in simple terms, the electronic wave functions of infinitely large systems with periodic boundary conditions
correspond to modulated plane waves, whereas a system with edges exhibits standing waves. 
Close to the edges, 
this different nature of the electronic wavefunctions leads to fluctuations in the the total electronic charge density. In metals, these fluctuations are known as Friedel oscillations, which decay algebraically with a wavelength $\lambda_\textrm{Friedel}=1/(2q_F)$, $q_F$ being the Fermi momentum \cite{Friedel}. In insulators instead, the charge deviations die out exponentially fast \cite{Prodan}. 
Consider for instance a finite one-dimensional atomic binary chain at half-filling: very close to the edges the electronic charge per unit cell $\rho_i$ starts deviating from its bulk value [c.f. Fig.~\ref{fig:fig1}]. 
This deviation can be  quantified by 
defining the total (excess) edge charge $Q_L$ 
as the sum of the local charge
deviations~\cite{Loss,loss2,loss3} $\Delta\rho_i=\rho_i-N_F$, where $N_F$ denotes the number of filled bands ($N_F=1$ for a the half-filled binary chain of Fig.~\ref{fig:fig1}), i.e.
\begin{align}\label{eq:QL}
Q_L:=\lim_{l\rightarrow\infty}\sum_{i=1}^{l} \Delta \rho_i.
\end{align}
In the equation above, the thermodynamic limit $l \rightarrow \infty$ explicitly accounts for a semi-infinite system, and we introduced the subindex $L$ to indicate that we refer to the left edge of the atomic chain. The edge charge, as defined in Eq.~\eqref{eq:QL}, can be numerically calculated up to $1/M$ corrections by considering a finite chain with a large number $M$ of unit cells,
and summing the local charge deviations in half of it. 

It turns out, however, that this edge effect can be exactly quantified using the geometric phase of the individual bulk electronic Bloch waves. In complete analogy with, e.g.,  two-dimensional time-reversal symmetry-broken topological insulators, where
the number of chiral edge channels 
is given as an integral of the Berry curvature,\cite{TKNN} 
the total edge charge in conventional insulators can be expressed
as an integral of the Berry potential. 
Such a relation has been shown in one-dimensional (1D) systems 
in Ref.~\onlinecite{Bardarson}. Specifically, the integral of the Berry potential over the one-dimensional Brillouin zone (BZ) yields an intra- and inter-cellular part, with the latter corresponding exactly to the (excess) edge charge $Q_L$, while the former quantifies the difference between the electronic contribution to the charge polarization and the edge charge itself ~\cite{Vanderbilt, kingsmith,Zak}. 
In this paper, we will exploit this relation to show that the excess charge can be formulated in terms of the topological invariants that classify insulating states in one-dimension protected by spatial symmetries.\cite{Bernevig,Chiu,Shiozaki,Lau3,Lau} In particular, for time-reversal symmetric systems this relation will be uncovered using the notion of partial Berry phases originally introduced by Fu and Kane.\cite{Fukane} 

The paper is organized as follows: In Sec.~\ref{sec:secedgecharge} we provide the 
derivation of the relation between the edge charge and the geometric (partial) Berry phase in 1D insulating systems. 
After reviewing the $\mathbb{Z}_2$ topology of 1D systems protected by point-group symmetries, 
we will show in Sec.~\ref{sec:sectopoinv} that the edge charge provides a natural probe for these free-fermion symmetry-protected topological (SPT) phases.  
Finally, we will draw our conclusions in Sec.~\ref{sec:secconc}.

\section{Edge charge of 1D systems}
\label{sec:secedgecharge}
In this section, we will demonstrate that the edge charge defined in Eq.~\eqref{eq:QL} for an atomic chain can be expressed as:
\begin{align}
\label{eq:bulkedge1}
Q_L&=-\frac{1}{2\pi}\sum_{n\leq N_F}\int_{-\pi}^\pi \mathrm{d}q\langle\Psi_n(q)|i\partial_q|\Psi_n(q)\rangle=-\dfrac{\gamma}{2\pi}.
\end{align}
Here, $|\Psi_n(q)\rangle$ denotes the entire Bloch wave with band index $n$ and crystal momentum $q$, and $\gamma$ is the Berry phase of the Bloch wavefunction $|\Psi_n(q)\rangle$. The inner-product is restricted to a single unit cell. In Appendix~\ref{sec:appendixa} we prove that the Berry phase is identical to the 
inter-cellular part of the Zak phase identified in Ref.~\onlinecite{Bardarson}. 
We stress that Eq.~\eqref{eq:bulkedge1} holds using the periodic gauge condition:
$|\Psi_m(q)\rangle=|\Psi_m(q+2\pi)\rangle$, where we put the lattice constant $a=1$. Throughout this paper we will always require that this periodicity condition will be obeyed.
\subsection{Notation}
Before 
providing the proof of
Eq.~\eqref{eq:bulkedge1}, we  introduce our notation. In the remainder we will limit ourselves to tight-binding models. This means that 
a generic 
Hamiltonian 
can be expressed as
\begin{align*}
\hat{H}&=\sum_{i,j}\sum_{\alpha,\beta}t_{j}^{\alpha,\beta}f^\dagger_{i,\alpha}f_{i+j,\beta},
\end{align*}
where $f^\dagger_{i,\alpha}$ is the creation operator corresponding to an electron in unit-cell $i$, 
and 
the index
$\alpha$, which runs from $1$ to $N$, refers to the electronic internal degrees of freedom. It
may therefore correspond to a spin, a sub-lattice or an orbital index. In the example of the binary chain introduced in Sec.~\ref{sec:secintro}, $\alpha$ corresponds to the sublattice index. 
The choice of the unit cell is fixed by the edge under consideration, see for example Fig.~\ref{fig:fig1} where the green rectangle denotes a preferred unit cell. To exploit the translation symmetry of the chain, we introduce the Fourier transformed creation and annihilation operators
\begin{align*}
f^\dagger_{q,\alpha}&:=\sum_{l=1}^{M}e^{iq l}f^\dagger_{l,\alpha}/\sqrt{M}.
\end{align*} 
Using these operators we can rewrite the Hamiltonian as
\begin{align*}
\hat{H}&=\sum_{q\in BZ}\hat{H}(q)=\sum_{\alpha,\beta}\sum_{q\in BZ}f^\dagger_{q,\alpha}\tilde{H}^{\alpha,\beta}(q)f_{q,\beta},
\end{align*}
where $\tilde{H}^{\alpha,\beta}(q)=\sum_jt_j^{\alpha,\beta}e^{iqj}$. We refer to $\hat{H}(q)$ as the second-quantized Hamiltonian, while $\tilde{H}(q)$ is its first quantized counterpart. 
We mention that we will use the same notation for 
other operators that we will introduce throughout this paper. 
We further denote the eigenstates of the first quantized Hamiltonian with $|\Psi_{n}(q)\rangle=[\Psi_{n,1}(q),\ldots,\Psi_{n,N}(q)]^T$, where $n=1,\ldots,N$ is the band index. 
The real-space wave function with crystal momentum $q$ and band index $n$ within a given unit cell is of course proportional to $|\Psi_n(q)\rangle$. 
\subsection{Derivation}
Having 
introduced
the notation, we now move on to derive Eq.~\eqref{eq:bulkedge1}. In Ref.~\onlinecite{Bardarson}, 
the correspondence between the excess charge and the inter-cellular part of the Zak phase has been found making use of Wanier orbitals. 
Here, we will take instead a different approach. Our proof consists of two parts, and relies on adiabatic deformation of an original Hamiltonian $\hat{H}_0$. In the first part, we show that Eq.~\eqref{eq:bulkedge1} holds for a simple tight-binding model described by the Hamiltonian $\hat{H}_0$. In the second part, we imagine that the tight-binding Hamiltonian $\hat{H}_0$ is adiabatically 
changed in time. 
Hence, we assume that we are 
provided with 
a one-parameter family of Hamiltonians $\hat{H}_\lambda$, where $\lambda$ denotes the parameter that varies in time. Then, we show that $\Delta Q_L:=Q_L(\lambda_f)-Q_L(\lambda_i)$ can be expressed as 
the difference of the Berry phases
$\left[\gamma(\lambda_i)-\gamma(\lambda_f)\right] /(2\pi)$. All together, this will prove the validity of Eq.~\eqref{eq:bulkedge1}. 

First, let us define $\hat{H}_0$ by considering an atomic chain where all electrons are completely localized within a unit cell and cannot hop to neighbouring unit cells, i.e. $t_j^{\alpha,\beta}=0$ for $j\neq0$. This ensures that 
for all momenta
$\tilde{H}_0(q)=\tilde{H}_0(0)$. Therefore, we find that the corresponding Bloch waves are identical: $|\Psi^0_n(q)\rangle=|\Psi^0_n(0)\rangle$. As a result,
the integrand on the right-hand side of Eq.~\eqref{eq:bulkedge1} vanishes. Since the edge charge for 
a system of perfectly localized electrons 
must identically vanish, we have proven that Eq.~\eqref{eq:bulkedge1} holds for $\hat{H}_0$.

\begin{figure}[t]
\centering
\includegraphics[width=.475\textwidth]{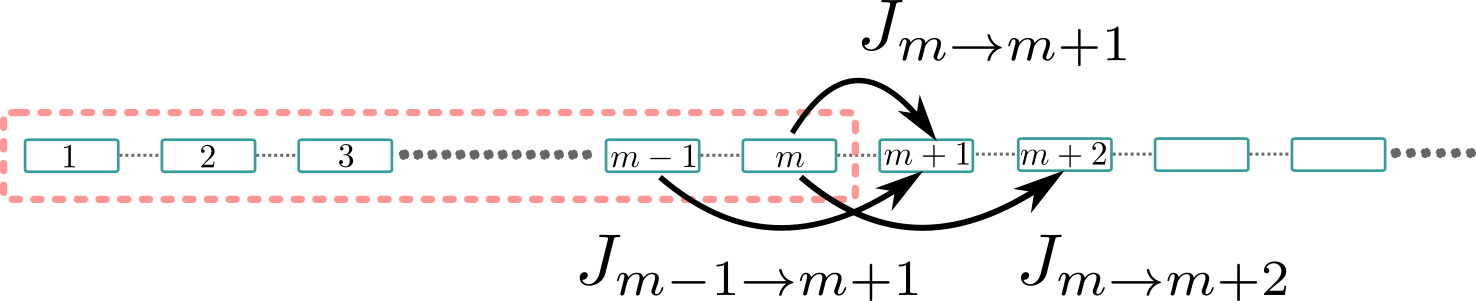}
\caption{One-dimensional chain. The green rectangles denote the unit-cells. The current that flows out of the red rectangular box is given by $J^\textrm{total}_{m\rightarrow m+1}=J_{m\rightarrow m+1}+J_{m-1\rightarrow m+1}+J_{m\rightarrow m+2}+\ldots$.}
\label{fig:fig2}
\end{figure}
Now let us turn to $\Delta Q_L$. By using Eq.~\eqref{eq:QL}, we can express the derivative of the edge charge as
\begin{align*}
\frac{dQ_L(\lambda(t))}{dt}&=\sum_{j=1}^\infty\frac{d\rho_j}{dt}\approxeq\sum_{j=1}^m\frac{d\rho_j}{dt},
\end{align*}
where $m$ is an arbitrarily large integer and we used that far away from the edges the charge per unit cell is constant. This allows us to write $\frac{d\rho_j}{dt}=0$ for $j\geq m$. By using the continuity equation, we then find
\begin{align}\label{eq:continuityeq}
\frac{dQ_L(\lambda(t))}{dt}&=-J^\textrm{total}_{m\rightarrow m+1}\nonumber\\
&=-\left(J_{m\rightarrow m+1}+J_{m\rightarrow m+2}+J_{m-1\rightarrow m+1}+\ldots\right)
\end{align}
In the equation above, $J^\textrm{total}_{m\rightarrow m+1}$ is the total current flowing through a wall put between unit-cells $m$ and $m+1$
[c.f. Fig.~\ref{fig:fig2}]. 
It
can be also written as the sum of the currents $J_{j\rightarrow k}$ flowing between two unit cells $j$ and $k$, with $j\leq m\leq k$. Note that this current does not capture any charge redistributions within the unit cell. These internal charge distributions are important for the charge polarization, but are irrelevant for the edge charge. The corresponding operator can be written as
\begin{align}\label{eq:currentop}
\hat{J}^\textrm{total}_{m\rightarrow m+1}(t)=\sum_{q\in BZ}f^\dagger_{q}\left[\nabla_q \tilde{H}_{\lambda(t)}(q)\right]f_{q}+\ldots
\end{align}
where the $\ldots$ indicate terms of the form $f^\dagger_q f_{q'}$  that couple different momentum states, {\it i.e.}  $q\neq q'$, and are completely irrelevant for a translational invariant bulk system. We refer the reader to Appendix~\ref{sec:appendixb}, for a derivation of Eq.~\eqref{eq:currentop}. Next, we consider the case in which $\lambda$ varies adiabatically slowly in time. This allows us to use the near-adiabatic approximation\cite{Thouless,Rigolin}, 
and express
the wavefunction at time $t$ 
as
\begin{align}\label{eq:nearadiabatic}
|\Psi_{n}(q,t)\rangle&=e^{if(q,t)}\left(|\Psi^{\lambda(t)}_{n}(q)\rangle+\right.\\
&\left.i\sum_{m\neq n}\frac{|\Psi^{\lambda(t)}_{m}(q)\rangle\langle\Psi^{\lambda(t)}_{m}(q))|\partial_t\Psi^{\lambda(t)}_{n}(q)\rangle}{E^{\lambda(t)}_m(q)-E^{\lambda(t)}_n(q)}\right)\nonumber.
\end{align}
Here, $|\Psi^\lambda_n(q)\rangle$ denotes the ``snapshot" Bloch wave function corresponding to $\hat{H}_\lambda(q)$, $E^\lambda_n(q)$ its instantaneous eigen energy, and $f$ is an arbitrary real-valued function. Combining Eqs.~\eqref{eq:continuityeq},~\eqref{eq:currentop}, and \eqref{eq:nearadiabatic}, we then obtain that the change in edge charge reads
\begin{widetext}
\begin{align*}
\Delta Q_L&=\frac{-i}{2\pi}\sum_{n\leq N_F}\sum_{m\neq n}\int_{t_i}^{{t_f}}\mathrm{d}t\int_{-\pi}^\pi\mathrm{d}q\langle \Psi_{n}^{\lambda(t)}(q)|\left[\nabla_q \tilde{H}_{\lambda(t)}(q)\right]|\Psi_{m}^{\lambda(t)}(q)\rangle\langle\Psi_{m}^{\lambda(t)}(q)|\partial_t\Psi_{n}^{\lambda(t)}(q)\rangle(E^{\lambda(t)}_m(q)-E^{\lambda(t)}_n(q))^{-1}+h.c.,
\end{align*}
\end{widetext}
where we have replaced the sum over $q$ by an integral. To make further progress, we eliminate the sum over $m$ by using 
\begin{align*}
\partial_q|\Psi_{n}^{\lambda}(q)\rangle&=\sum_{m\neq n}\frac{\langle\Psi_{m}^\lambda(q)|\left[\nabla_q \tilde{H}_\lambda(q)\right]|\Psi_{n}^\lambda(q)\rangle}{E_n^\lambda(q)-E_m^\lambda(q)}|\Psi_{m}^\lambda(q)\rangle+\nonumber\\
&i g(q)|\Psi_{n}^\lambda(q)\rangle,
\end{align*}
where $g$ is an  arbitrary real-valued function that does not contribute to the integral. Hence, we find
\begin{align}\label{eq:DQL}
\Delta Q_L&=\frac{i}{2\pi}\sum_{n\leq N_F}\int_{t_i}^{{t_f}}\mathrm{d}t \int \mathrm{d}q \langle\partial_q\Psi_{n}^{\lambda(t)}(q)|\partial_t \Psi_{n}^{\lambda(t)}(q)\rangle+h.c.
\end{align}
Using Stokes theorem we can rewrite 
the r.h.s. of the equation above 
as a line integral. By further imposing the periodic gauge for the wave function $|\Psi^\lambda_m(q)\rangle=|\Psi^\lambda_m(q+2\pi)\rangle$, Eq.~\eqref{eq:DQL} assumes the following form
\begin{align*}
\Delta Q_L&=\frac{-i}{2\pi}\sum_{n\leq N_F}\int\mathrm{d}q\langle\Psi_{n}^{\lambda_f}(q)|\partial_q\Psi_{n}^{\lambda_f}(q)\rangle\nonumber\\
&+\frac{i}{2\pi}\int\mathrm{d}q\langle\Psi_{n}^{\lambda_i}(q)|\partial_q\Psi_{n}^{\lambda_i}(q)\rangle=\frac{1}{2\pi}\left(\gamma(\lambda_i)-\gamma(\lambda_f)\right).
\end{align*}
With this, 
we have shown that the edge charge is indeed given by Eq.~\eqref{eq:bulkedge1}. We note that the Berry phase $\gamma$ can be conveniently expressed in terms of the trace of the non-Abelian Berry potential $\mathcal{A}_{m,n}(q)=\langle\Psi_{m}(q)|i\partial_q|\Psi_{n}(q)\rangle$, with $m,n=1,\ldots N_F$:
\begin{align}\label{eq:gammanonABEL}
\gamma=\int_{-\pi}^\pi\mathrm{d}q\Tr{\mathcal{A}(q)}.
\end{align}
This expression is invariant under a $U(N_F)$ gauge transformation $|\Psi_m(q)\rangle\rightarrow \mathcal{U}^{m,n}(q)|\Psi_n(q)\rangle$, for which Eq.~\eqref{eq:gammanonABEL} is transformed accordingly to
\begin{align}\label{eq:GT1}
\Tr{\mathcal{A}(q)}\rightarrow& \Tr{\mathcal{A}(q)}+i\Tr{\mathcal{U}^\dagger(q)\partial_q\mathcal{U}(q)}
\end{align}
Since, $\mathcal{U}$ is a unitary matrix, we find
\begin{align}\label{eq:GT2}
i\Tr{\mathcal{U}^\dagger(q)\partial_q\mathcal{U}(q)}=i\partial_q\log{\det{\mathcal{U}(q)}}
\end{align}
With this, it follows that $\gamma\rightarrow\gamma+2\pi j$, with $j$ the winding number $W(\mathcal{U})$ of the determinant of $\mathcal{U}(q)$, which is given by
\begin{align*}
W(\mathcal{U})&=\frac{i}{2\pi}\int_{-\pi}^\pi\mathrm{d}q\frac{d}{dq}\log{\det{\mathcal{U}(q)}}.
\end{align*}
Moreover we point out that the edge charges at the two opposite edges of a one-dimensional chain must compensate each other modulo $1$. Note that this is only true if the chain consists of an integer number of unit-cells. Hence, we can generally write 
\begin{align}\label{eq:bulkedge2}
Q&=\pm\frac{\gamma}{2\pi},
\end{align}
where $+$($-$) refers to a right (left) edge. We stress that Eq.~\eqref{eq:bulkedge2} is completely generic, and can be used to calculate the edge charge for any 1D crystalline insulator.

\begin{figure}[b]
\centering
\includegraphics[width=.475\textwidth]{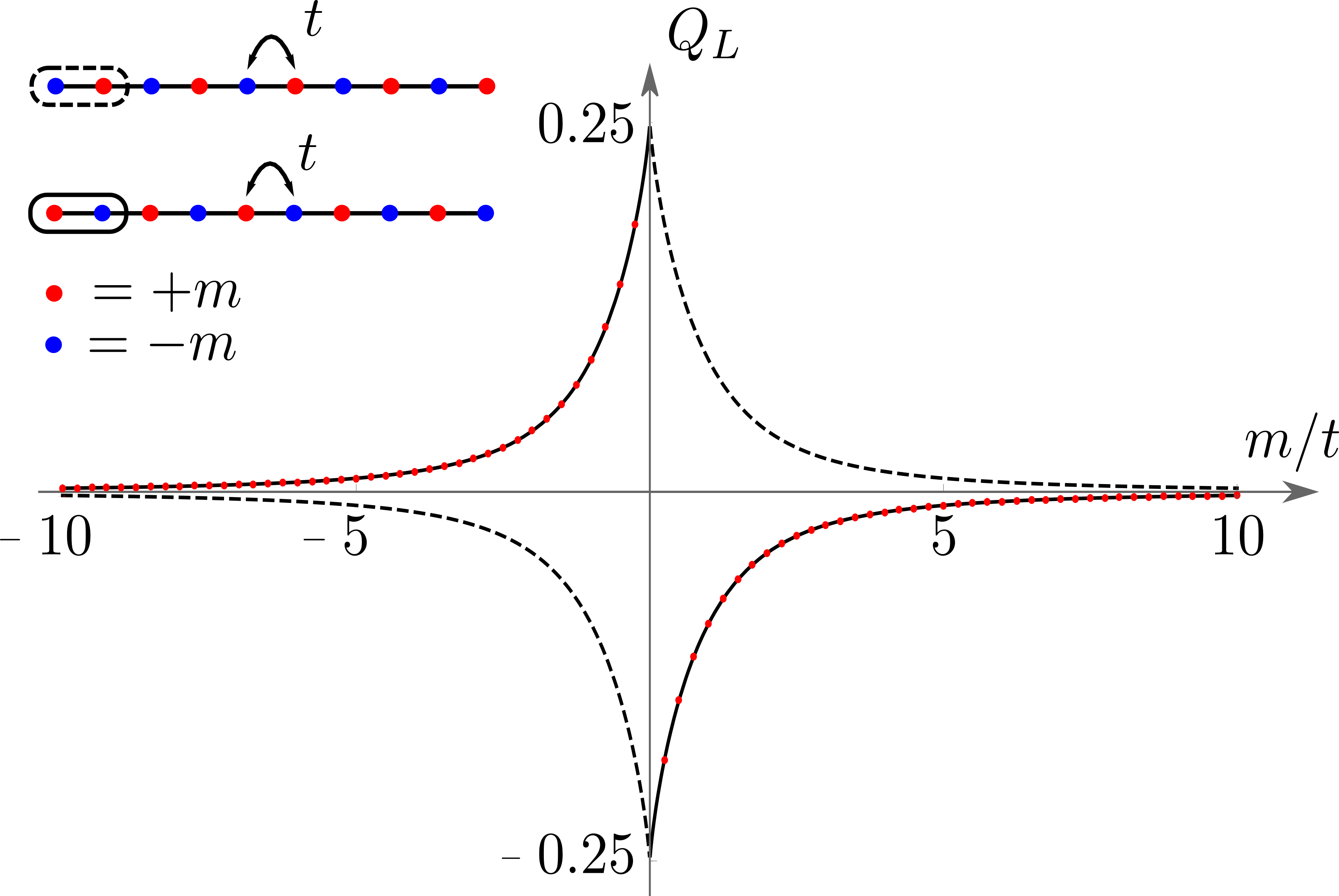}
\caption{The edge charge $Q_L$ for the binary chain. In the inset we display the two possible terminations. The solid and dashed line correspond to the Berry phase result for  the red and blue termination, respectively. The red dots denote the values for the edge charge for the red termination, which are obtained by numerical diagonalization for a chain of 200 unit cells at half-filling, using Eq.~\eqref{eq:QL}.}
\label{fig:fig3}
\end{figure}
Let us now take into account the binary chain introduced above to illustrate this result. First we have to choose a termination, which fixes the preferred unit cell. The binary chain can only be terminated in two ways, either with a blue site or with a red site. For the blue (red) termination the preferred unit cell is denoted with a solid (dashed) box in the inset of Fig.~\ref{fig:fig3}. The corresponding Fourier transformed Hamiltonians $\tilde{H}^\textrm{red}(q)$ and $\tilde{H}^\textrm{blue}(q)$ are given by
\begin{align*}
\tilde{H}^\textrm{red}(q)&=\begin{pmatrix}
m&t(1+e^{-i q})\\
t(1+e^{i q})&-m
\end{pmatrix},
\end{align*}
and
\begin{align*}
\tilde{H}^\textrm{blue}(q)&=\begin{pmatrix}
-m&t(1+e^{-i q})\\
t(1+e^{i q})&m
\end{pmatrix}.
\end{align*}
At half-filling the left edge charges corresponding to the blue and red termination are plotted as a function of $m/t$ in Fig.~\ref{fig:fig3}. Note that both vanish in the limit $|m/t|\rightarrow\infty$. This is expected, as it corresponds to the
atomic limit in which the hopping amplitude goes to zero. Moreover, from Fig.~\ref{fig:fig3} we immediately notice that $Q^\textrm{blue}=-Q^\textrm{red}$.  This follows from the fact that the red and blue termination are related by inversion-symmetry. The same symmetry 
 yields
 a
 $\pi$ jump 
 in the Berry phase
 at $m=0$. We will discuss this in more detail in Sec.~\ref{sec:sectopoinv}.  The red dots in Fig.~\ref{fig:fig3} are obtained by numerically calculating the left edge charge for a chain with $200$ unit cells using Eq.~\ref{eq:QL}.
\subsection{Termination dependence}
\begin{figure}[t]
\centering
\includegraphics[width=.475\textwidth]{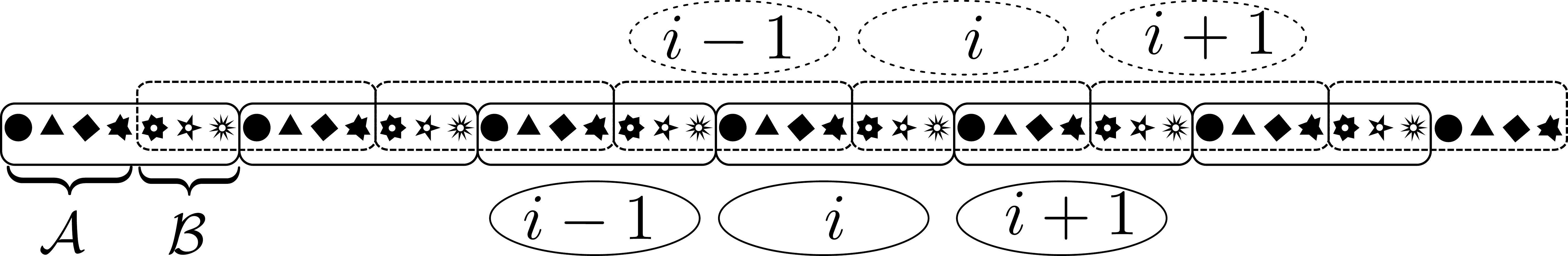}
\caption{One-dimensional chain with two different unit-cells solid and dashed. The solid (dashed) ellipses denote the corresponding labelling of the unit cells. Sites within partition $\mathcal{A}$ are completely black, whereas the sites in partition $\mathcal{B}$ have a white center.}
\label{fig:fig4}
\end{figure}
We next investigate how the edge charges for two different terminations are related. For this purpose we consider a generic tight-binding model, see Fig.~\ref{fig:fig4}, for a sketch. In addition, we have depicted solid and dashed unit cells, which we refer to as unit cells $1$ and $2$,  respectively. Next, let us analyze how the corresponding creation operators, $f^\dagger_{i,\alpha,1}$ and $f^\dagger_{i,\alpha,2}$ are related. To this end, we partition the unit-cell into two parts, called $\mathcal{A}$ and $\mathcal{B}$, see Fig.~\ref{fig:fig4}. We relabel the creation operators in partition $\mathcal{A}$($\mathcal{B}$) as $a^\dagger_{i,\alpha,1}$($b^\dagger_{i,\alpha,1}$ ) and  $a^\dagger_{i,\alpha,2}$($b^\dagger_{i,\alpha,2}$). Then, it immediately follows that the creation operators for the two different unit cells are related by
\begin{align*}
a^\dagger_{i,\alpha,1}&=a^\dagger_{i-1,\alpha,2},\\
b^\dagger_{i,\alpha,1}&=b^\dagger_{i,\alpha,2}.
\end{align*}
By performing a Fourier transformation and writing $f^\dagger_{q,1}=(a^\dagger_{q,1},b^\dagger_{q,1})$ and $f^\dagger_{q,2}=(b^\dagger_{q,2},a^\dagger_{q,2})$,  we find that
\begin{align*}
f^\dagger_{q,1}=f^\dagger_{q,2}\tilde{U}(q),
\end{align*}
where the matrix $\tilde{U}(q)$ is given by
\begin{align*}
\tilde{U}(q)&=\begin{pmatrix}
0&1\\
e^{iq}&0
\end{pmatrix}.
\end{align*}
From this, it follows that the Bloch waves for the two unit-cells are related by $\tilde{U}(q)|\Psi^1_n(q)\rangle=|\Psi^2_n(q)\rangle$. 
This, in turns, implies 
that $\gamma^1=\gamma^2-2\pi\rho_\mathcal{B}$, where $\rho_\mathcal{B}$ denotes the total charge in the $\mathcal{B}$ partition. 
The knowledge of the Berry phase for one unit cell and of the charge distribution within that particular unit-cell then allows to
compute the Berry phase for all possible unit cells. For the binary chain we have explicitly verified 
this relation for the edge charges considering the blue and red terminations. 

\begin{figure}[t]
\centering
\includegraphics[width=.475\textwidth]{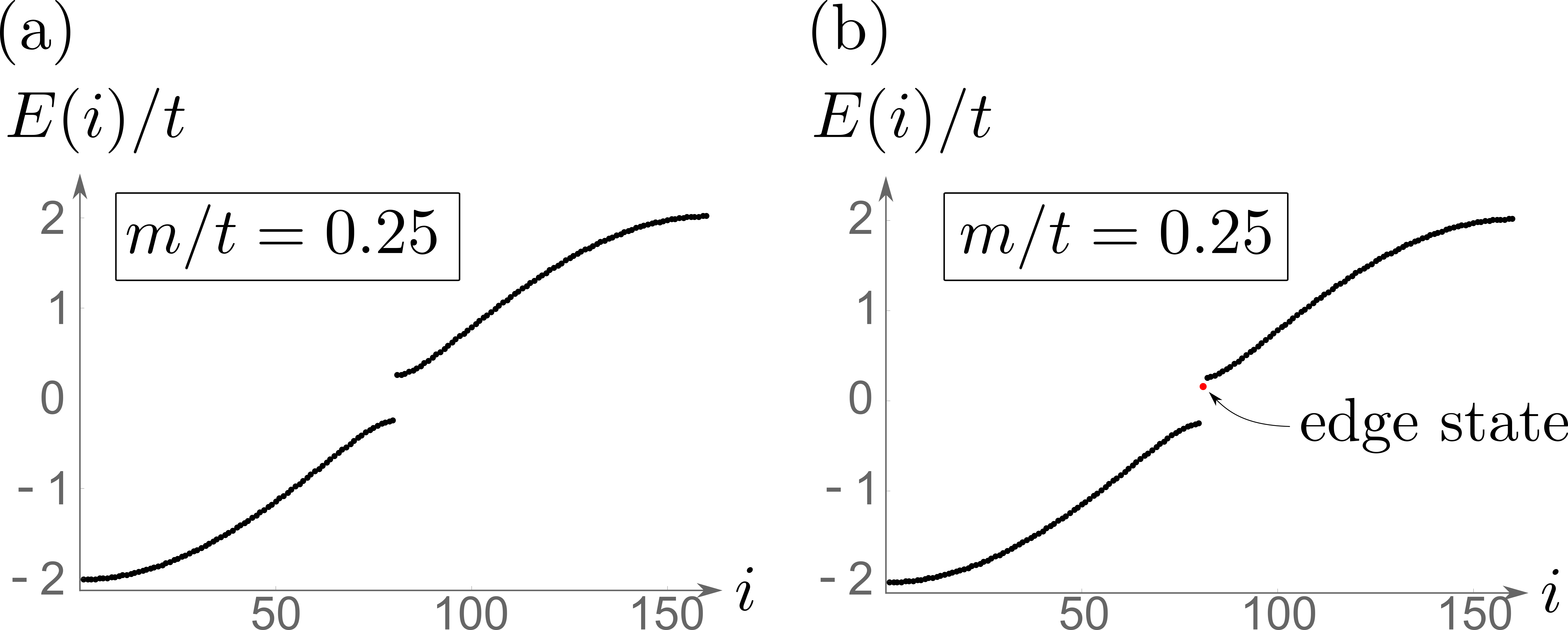}
\caption{Two spectra for the binary chain with $80$ unit cells, with different edge potentials. In (a) all red sites are at on-site energy $+m$, whereas in (b) the first red site is at on-site energy $-m$. In (b) the edge potential gives rise to an edge state.}
\label{fig:fig5}
\end{figure}

Finally, let us address the bulk nature of the edge charge. Since the Berry phases are only well-defined up to integer multiples of $2\pi$, we can only predict the fractional part of the edge charge. 
We emphasize that 
this is not a limitation of the Berry phase approach, but an intrinsic property of the edge charge. 
Specifically, 
the integer part of the edge charge depends on microscopic details of the termination as well as on the Fermi level $E_F$. 
For instance, 
the edge spectrum may host edge states depending on the details of the edge potential. The occupancy of these states, which is controlled by $E_F$, changes the edge charge by $\pm 1$. To illustrate this, we consider the binary chain terminated with the red site. If we put the first site at an on-site energy $-m$ instead of $+m$, we find that the spectrum exhibits an in-gap state, see Fig.~\ref{fig:fig5}. Eq.~\eqref{eq:QL} gives $Q_L=(-0.18) 0.82 $ if this state is (un)-occupied, which agrees with the result of Eq.~\eqref{eq:bulkedge1} modulo an integer [see Fig.~3].
This result corroborates the fact that only
the fractional part of the excess charge is a bulk quantity, and can be thus expressed as a geometric phase. 

\subsection{Time-reversal symmetry}
So far we have considered the edge charge without explicitly invoking time-reversal symmetry.
For spin one-half fermions, 
Kramer's theorem guarantees that every state  is necessarily doubly degenerate. In particular, this applies to the in-gap edge states. Using the result above, this also implies that the edge charge can only change by multiples of $2$ upon changing the Fermi level, thereby suggesting that 
the relation between excess charges and quantum-mechanical geometric phases can be refined when explicitly accounting for time-reversal symmetry. 

\begin{figure}[t]
\centering
\includegraphics[width=.475\textwidth]{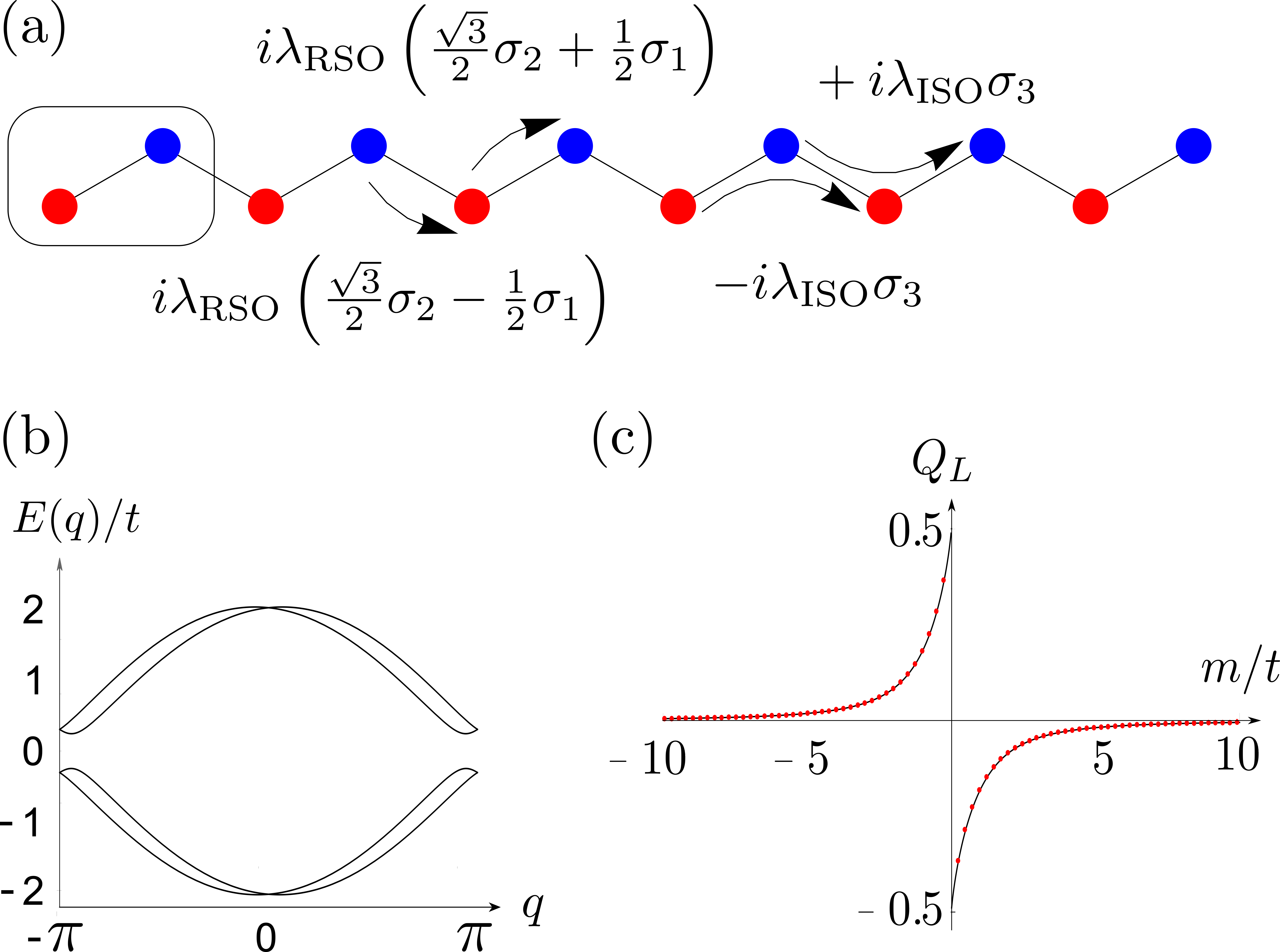}
\caption{Spinfull binary chain. (a) Spin-orbit coupling terms are schematically depicted with arrows. (b) Typical band structure for a system consisting of spin one-half electrons in the presence of time-reversal symmetry. Note the Kramer's degeneracies at $q=0$ and $q=\pi$.  (c) The edge charge for the spin-full binary chain. The solid line corresponds to the partial Berry phase result, whereas the red dots are obtained by numerical diagonalization for an open chain of 200 unit cells. Here, we have chosen $\lambda_\textrm{ISO}/t=\lambda_\textrm{RSO}/t=1/10$.}
\label{fig:fig6}
\end{figure}

We start out by considering that time-reversal symmetry imposes the following constraints: 
\begin{align*}
\hat{H}(q)=\hat{T}\hat{H}(-q)\hat{T}^{-1}\quad \textrm{and } \hat{T}^2=-1, 
\end{align*}
with $\hat{T}$ the anti-unitary time-reversal symmetry operator. 
This constraint ensures that the band structure consists of pairs of bands, which touch at the time-reversal invariant momenta, see Fig.~\ref{fig:fig6}(b). We label the different pairs by $n=1,\ldots,N_F/2$. Moreover, for a given pair with index $n$ and momentum $q$, we refer to the two states as $|\Psi^I_{n}(q)\rangle$ and $|\Psi^{II}_{n}(q)\rangle$. 
Let us for the moment assume that we have found a
smooth time-reversal symmetric gauge, i.e.
\begin{align}\label{eq:TRSgauge}
|\Psi^{II}_{n}(q)\rangle=\tilde{T}|\Psi^{I}_{n}(-q)\rangle.
\end{align}
Where $\tilde{T}$ is the first-quantized anti-unitary operator corresponding to $\hat{T}$. Using this decomposition, we can rewrite Eq.~\eqref{eq:DQL} as
\begin{widetext}
\begin{align*}
\Delta Q_L&=\frac{-i}{2\pi}\sum_{n\leq N_F/2}\int_{t_i}^{{t_f}}\mathrm{d}t\int_{-\pi}^\pi\mathrm{d}q \langle\partial_q\Psi^{I,\lambda(t)}_{n}(q)|\partial_t \Psi^{I,\lambda(t)}_{n}(q)\rangle-\frac{i}{2\pi}\sum_{n\leq N_F/2}\int_{t_i}^{{t_f}}\mathrm{d}t\int_{-\pi}^\pi\mathrm{d}q \langle\partial_q\Psi^{II,\lambda(t)}_{n}(q)|\partial_t \Psi^{II,\lambda(t)}_{n}(q)\rangle+h.c.\nonumber\\
&=\frac{-i}{\pi}\sum_{n\leq N_F/2}\int_{t_i}^{{t_f}}\mathrm{d}t\int_{-\pi}^\pi\mathrm{d}q \langle\partial_q\Psi^{I,\lambda(t)}_{n}(q)|\partial_t \Psi^{I,\lambda(t)}_{n}(q)\rangle+h.c.=\frac{1}{\pi}(\gamma^I(\lambda_i)-\gamma^I(\lambda_f)),
\end{align*}
\end{widetext}
where in the final line we have employed Stokes' theorem to rewrite the surface integral as a contour integral, and introduced the partial Berry phase\cite{Fukane} $\gamma^I$ which is defined modulo $2 \pi$. This confirms that the edge charge in time-reversal symmetric systems is indeed well-defined modulo $2$, and expressed in terms of the partial Berry phase by
\begin{align}\label{eq:bulkedgetrs1}
Q&=\pm\sum_{n\leq N_F/2}\frac{1}{\pi}\int_{-\pi}^\pi\mathrm{d}q\langle\Psi^I_n(q)|i\partial_q|\Psi^I_n(q)\rangle=\pm\gamma^I/\pi.
\end{align}
In the equation above, the $+$($-$) refers again to the right (left) edge. 

Since 
it is not always an easy task to find a smooth gauge in time-reversal symmetric systems, we next wish to find a formulation of Eq.~\eqref{eq:bulkedgetrs1}, which is invariant under an arbitrary gauge transformation. First, let us point out that the time-reversal symmetric gauge Eq.~\eqref{eq:TRSgauge}, assures that
\begin{align*}
\langle \Psi^I_n(q)|i\partial_q|\Psi^I_n(q)\rangle=\langle \Psi^{II}_n(-q)|i\partial_{-q}|\Psi^{II}_n(-q)\rangle.
\end{align*}
This allows us to express the partial Berry phase as an integral of the trace of the non-Abelian Berry potential over half the Brillouin zone
\begin{align}\label{eq:partialgamma1}
\gamma^I&=\int_0^\pi\mathrm{d}q\Tr{\mathcal{A}(q)}.
\end{align}
Next, we introduce the sewing matrix $\mathcal{S}_{\tilde{T}}(q)$ whose entries are given by
\begin{align*}
\left[\mathcal{S}_{\tilde{T}}(q)\right]^{m,n}&=\langle\Psi_m(-q)|\tilde{T}|\Psi_n(q)\rangle.
\end{align*}
The sewing matrix is anti-symmetric at the time-reversal invariant momenta $q=0,\pi$, and as such 
can be characterized by 
its Pfaffian. As long as Eq.~\eqref{eq:TRSgauge} is obeyed, we find that 
\begin{align}\label{eq:Pf}
\Pf{\mathcal{S}_{\tilde{T}}(\pi)}/\Pf{\mathcal{S}_{\tilde{T}}(0)}&=1.
\end{align}
Since, the log of $1$ is zero, we can freely add Eq.~\eqref{eq:Pf} to Eq.~\eqref{eq:partialgamma1},
\begin{align}\label{eq:partialgamma2}
\gamma^I&=\int_0^\pi\mathrm{d}q\Tr{\mathcal{A}(q)}+i\log{\left(\Pf{\mathcal{S}_{\tilde{T}}(\pi)}/\Pf{\mathcal{S}_{\tilde{T}}(0)}\right)}
\end{align}
The advantage of this expression is that it is invariant under an arbitrary $U(N_F)$ gauge transformation. Using Eqs.~\eqref{eq:GT1} and  \eqref{eq:GT2}, we find that under a gauge transformation the first term in the r.h.s. of the equation above changes by 
\begin{align*}
i\log{\det{\mathcal{U}(\pi)}/\det{\mathcal{U}(0)}}.
\end{align*}
The sewing matrices instead transform as
\begin{align*}
\mathcal{S}_{\tilde{T}}(0)&\rightarrow \mathcal{U}^\dagger(0)\mathcal{S}_{\tilde{T}}(0)\mathcal{U}(0)^*\\
\mathcal{S}_{\tilde{T}}(\pi)&\rightarrow\mathcal{U}(\pi)^\dagger\mathcal{S}_{\tilde{T}}(\pi)\mathcal{U}(\pi)^*.
\end{align*}
Using the fact that $\Pf{X A X^T}=\Pf{A}\det{X}$, we find that the second term in the r.h.s. of Eq.~\eqref{eq:partialgamma2} changes by 
\begin{align*}
i\log\left[\det{\mathcal{U}^\dagger(\pi)}/\det{\mathcal{U}^\dagger(0)}\right]&=-i\log\left[\det{\mathcal{U}(\pi)}/\det{\mathcal{U}(0)}\right].
\end{align*}
Hence, this proves that the right hand side of Eq.~\eqref{eq:partialgamma2} is gauge invariant and does not necessitate Eq.~\eqref{eq:TRSgauge} to be fulfilled.

To numerically confirm these results, 
let us consider a spinful version of the binary chain. Assuming that the orbitals are real, we find that the time-reversal operator $\tilde{T}=\mathbb{I}\otimes i\sigma_2\mathcal{K}$, where the identity acts on the orbital and sub-lattice degrees of freedom, the second Pauli matrix $\sigma_2$ on the spin, and $\mathcal{K}$ corresponds to complex conjugation. In addition to the spinless part, we add both intrinsic and Rashba spin-orbit coupling. The former manifests itself through complex next-nearest neighbour hoppings, see Fig.~6(a). The corresponding Hamiltonian reads
\begin{align*}
\tilde{H}_\textrm{ISO}(q)&=\lambda_\textrm{ISO}\begin{pmatrix}
2\sin(q)&0\\
0&-2\sin(q)
\end{pmatrix}\otimes\sigma_3
\end{align*} 
The Rashba spin-orbit coupling is given instead by
\begin{align*}
\tilde{H}_\textrm{RSO}(q)&=i\lambda_\textrm{RSO}\begin{pmatrix}
0&1\\
-1&0
\end{pmatrix}\otimes\left(\frac{\sqrt{3}}{2}\sigma_2+\frac{1}{2}\sigma_1\right)-\\
&i\lambda_\textrm{RSO}\begin{pmatrix}
0&e^{-iq}\\
-e^{iq}&0
\end{pmatrix}\otimes\left(\frac{\sqrt{3}}{2}\sigma_2-\frac{1}{2}\sigma_1\right).
\end{align*}
It is easily verified that $\tilde{H}_\textrm{ISO}(q)=\tilde{T}\tilde{H}_\textrm{ISO}(-q)\tilde{T}$ and $\tilde{H}_\textrm{RSO}(q)=\tilde{T}\tilde{H}_\textrm{RSO}(-q)\tilde{T}$. The corresponding band structure is depicted in Fig.~6(b), where we used $m/t=0.25$ and $\lambda_\textrm{RSO}/t=\lambda_\textrm{ISO}/t=1/10$. Note that apart from the time-reversal invariant momenta the bands are completely spin-split. The edge charge is calculated using the partial Berry phase $\gamma^I$ for various values of $m/t$, see Fig.~6(c). The red dots denote the values obtained for the edge charges by numerical diagonalization for a chain of $200$ unit-cells at half-filling. And indeed we find that the partial Berry phase correctly predicts the edge charges mod $2$. 

\subsection{Numerical considerations}
To compute a (partial) Berry phase one should find a smooth gauge. In practice, this requires 
 to impose a certain gauge-fixing condition. For example, one might fix the gauge by requiring that the wave function is strictly real and positive at a certain site. However, such a gauge-fixing condition becomes ill-defined if the wave-function vanishes at this site. Fortunately, there is an easier method to calculate (partial) Berry phases, see for example Ref.~\onlinecite{kingsmith}. Here, we briefly discuss these methods.

Suppose that $|\Psi_{n}(q)\rangle$ is a smooth gauge. Then we can define the $N_F\times N_F$ overlap matrix 
\begin{align*}
\mathcal{S}_{m,n}(q_1,q_2):=\langle\Psi_m(q_1)|\Psi_{n}(q_2)\rangle.
\end{align*}
This yields
\begin{align*}
\mathcal{S}(q,q+\epsilon)=e^{-i\epsilon \mathcal{A}(q)}+\mathcal{O}(\epsilon^2).
\end{align*}
Let now $q_j:=j2\pi/N$, with $j=0,1,\ldots,N$, be a discretization of the 1D BZ. If we use that $\det\left[\mathcal{S}(q,q+\epsilon)\right]=e^{-i\epsilon\Tr\left[\mathcal{A}(q)\right]}$, we find
\begin{align*}
\lim_{N\rightarrow\infty}\det\left[\prod_{i=0}^{N-1}\mathcal{S}(q_i,q_{i+1})\right]=e^{-i\gamma}.
\end{align*}
Note that the l.h.s. of the equation above 
is invariant under an arbitrary $U(N_F)$ gauge transformation $|\Psi_{m}(q_i)\rangle\rightarrow \mathcal{U}^{m,n}(q_i)|\Psi_{n}(q_i)\rangle$, as long as the periodicity $|\Psi_{m}(
q_0)\rangle=|\Psi_{m}(q_N)\rangle$ is respected. This removes the necessity to find a smooth gauge. More importantly, it provides a practical method to calculate the Berry phase. 

Similarly, one can calculate the partial Berry phase $\gamma^I$. Following the same steps as above we find
\begin{align*}
\lim_{N\rightarrow\infty}\det\left[\prod_{i=0}^{N-1}\mathcal{S}(\tilde{q}_i,\tilde{q}_{i+1})\right]\cdot\frac{\Pf{\mathcal{S}_{\tilde{T}}(\pi)}}{\Pf{\mathcal{S}_{\tilde{T}}(0)}}=e^{-i\gamma^I},
\end{align*}
where we introduced the mesh $\tilde{q}_j=j\pi/N$. 
The l.h.s. of the equation above provides a practical method to calculate the partial Berry phase.

\section{Edge charge as a probe of band structure topology}
\label{sec:sectopoinv}
In this section we will discuss the $\mathbb{Z}_2$-classification of 1D crystalline insulators that are invariant under spatial symmetries interchanging the left and right edges of a chain ~\cite{Niu,Zaksym,Kohn,Fang,Alexandrinata}, and show that the edge charge can be used to probe the corresponding crystalline topological invariants.  
We will restrict our analysis to inversion, two-fold rotation, and mirror symmetry, 
and, as before, we will first not explicitly invoke the fermionic time-reversal symmetry. 

\begin{figure}[t]
\centering
\includegraphics[width=.475\textwidth]{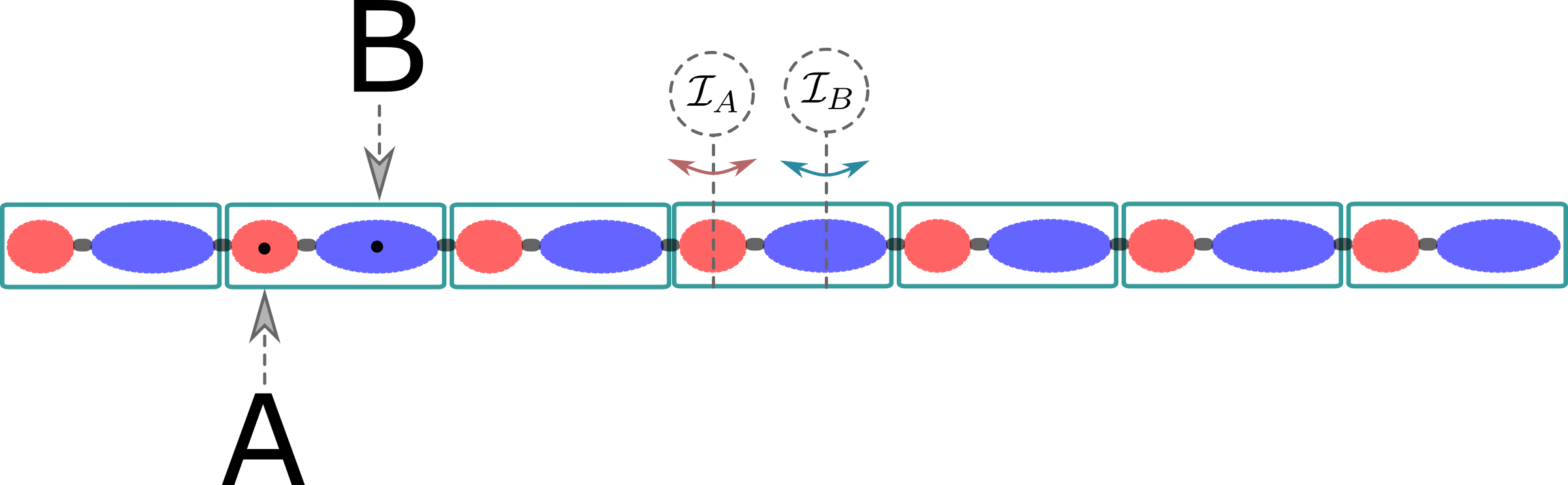}
\caption{Generic inversion-symmetric crystal. The symmetry centers are denoted with {
\fontfamily{cmss}\selectfont
\textbf{A}
} and  {
\fontfamily{cmss}\selectfont
\textbf{B}
}.}
\label{fig:fig7}
\end{figure}
When one considers a point-group symmetry in a crystal,
one should always specify the symmetry-center. In particular, an inversion-symmetric one-dimensional chain 
exhibits
two points of inversion per unit-cell, to which we refer as $A$ and $B$, see Fig.~\ref{fig:fig7}. Without loss of generality, we consider the case in which $B$ sits to the right of $A$ within the unit cell. Now let us consider how the canonical creation operators $f^\dagger_{i,\alpha}$ transform under inversion. To be as general as possible, we allow for a non-inversion symmetric unit-cell. We partition this unit cell into two parts, called $\mathcal{A}$ and $\mathcal{B}$, centered around the inversion points $A$ and $B$, respectively, see Fig~\ref{fig:fig7}. We denote the creation operators corresponding to orbitals and spin in part $\mathcal{A}$ ($\mathcal{B}$) with $a^\dagger_{i,\alpha}$ ($b^\dagger_{i,\alpha}$), such that $f^\dagger_i=(a^\dagger_{i,1},\ldots,a^\dagger_{i,n_\mathcal{A}},b^\dagger_{i,1},\ldots,b^\dagger_{i,n_\mathcal{B}})$. Next, we consider how the electrons transform under inversion. Inspection of Fig.~7 shows that under inversion through $A$ ($B$) the electrons in partition $\mathcal{A}$ within unit cell $i$ transform to partition $\mathcal{A}$ in unit cell $-i$ ($-i+1$). Hence, if $\hat{I}_A$ ($\hat{I}_B$) denotes the corresponding inversion operator, then we find
\begin{align*}
\hat{I}_A a^\dagger_{i,\alpha}\hat{I}_A^{-1}&=a^\dagger_{-i,\beta}\tilde{I}_\mathcal{A}^{\beta,\alpha},\textrm{ and}\\
\hat{I}_B a^\dagger_{i,\alpha}\hat{I}_B^{-1}&=a^\dagger_{-i+1,\beta}\tilde{I}_\mathcal{A}^{\beta,\alpha}.
\end{align*}
Similarly, we find that the electrons in partition $\mathcal{B}$ and unit cell $i$ are sent to partition $\mathcal{B}$ within the ($-i-1$)th ($-i$th) unit cell, upon inverting through point $A$ ($B$). Hence, we find
\begin{align*}
\hat{I}_A b^\dagger_{i,\alpha}\hat{I}_A^{-1}&=b^\dagger_{-i-1,\beta}\tilde{I}_\mathcal{B}^{\beta,\alpha},\textrm{ and}\\
\hat{I}_B b^\dagger_{i,\alpha}\hat{I}_B^{-1}&=b^\dagger_{-i,\beta}\tilde{I}_\mathcal{B}^{\beta,\alpha}.
\end{align*}
Next, we apply a Fourier transformation, and combine the above equations by writing $\hat{I}_Af^\dagger_{q,\alpha}\hat{I}_A^{-1}=f^\dagger_{-q,\beta}\tilde{I}^{\beta,\alpha}_A(q)$, with
\begin{align*}
\tilde{I}_A(q)&=\begin{pmatrix}
\tilde{I}_\mathcal{A}&0\\
0&e^{-iq}\tilde{I}_\mathcal{B}
\end{pmatrix}.
\end{align*}
Similarly, we find $\hat{I}_Bf^\dagger_{q,\alpha}\hat{I}_B^{-1}=f^\dagger_{-q,\beta}\tilde{I}^{\beta,\alpha}_B(q)$, with
\begin{align*}
\tilde{I}_B(q)&=\begin{pmatrix}
e^{iq}\tilde{I}_\mathcal{A}&0\\
0&\tilde{I}_\mathcal{B}
\end{pmatrix}.
\end{align*}
Since inversion symmetry squares to one, we find 
\begin{align}\label{eq:Isquared}
\tilde{I}_A(q)\tilde{I}_A(-q)=1=\tilde{I}_B(q)\tilde{I}_B(-q).
\end{align}
In addition, we like to point out that  $\tilde{I}_A(q)=e^{-iq}\tilde{I}_B(q)$. Hence, all properties of $\tilde{I}_B(q)$ can be obtained from $\tilde{I}_A(q)$. Therefore we limit ourselves in the following to $\tilde{I}_A(q)$.

The fact that the Hamiltonian is inversion symmetric ensures that the Fourier transformed Hamiltonians $\hat{H}(q)$ $\hat{H}(-q)$ are related by
\begin{align*}
\hat{I}_A\hat{H}(q)\hat{I}^{-1}_A=\hat{H}(-q),
\end{align*}
which reduces to 
\begin{align}\label{eq:inversionH}
\tilde{I}_A(q)\tilde{H}(q)\tilde{I}^{-1}_A(q)=\tilde{H}(-q),
\end{align}
using the first-quantized Hamiltonians. Armed with this structure, one can consider the sewing matrix $\mathcal{S}_{\tilde{I}_A}(q)$, given by
\begin{align}
\left[\mathcal{S}_{\tilde{I}_A}(q)\right]^{m,n}&=\langle\Psi_m(-q)|\tilde{I}_A(q)|\Psi_n(q)\rangle.
\end{align}
Here, Eq.~\eqref{eq:inversionH} guarantees that $\mathcal{S}_{\tilde{I}_A}(q)$ is a unitary matrix. Together with the fact that $\mathcal{S}_{\tilde{I}_A}(q)$ is $2\pi$ periodic, we can consider, assuming a smooth gauge for the Bloch wavefunctions, the winding number $W(\mathcal{S}_{\tilde{I}_A})$ of the determinant of the sewing matrix
\begin{align}\label{eq:windingSIA}
W(\mathcal{S}_{\tilde{I}_A}):=\frac{i}{2\pi}\int_{-\pi}^\pi\mathrm{d}q\frac{d}{dq}\log{\det{\mathcal{S}_{\tilde{I}_A}(q)}}\in\mathbb{Z}
\end{align}
Naively, one might believe that $W(\mathcal{S}_{\tilde{I}_A})$ yields a $\mathbb{Z}$ classification of $1D$ inversion-symmetric insulators. However, this winding number is not gauge-invariant.  Under a gauge transformation $|\Psi_m(q)\rangle\rightarrow \mathcal{U}^{m,n}(q)|\Psi_n(q)\rangle$, we find that $\det{\mathcal{S}_{\tilde{I}_A}(q)}\rightarrow\det{\mathcal{U}^\dagger(-q)}\det{\mathcal{S}_{\tilde{I}_A}(q)}\det{\tilde{\mathcal{U}}(q)}$. It follows that $W(\mathcal{S}_{\tilde{I}_A})\rightarrow W(\mathcal{S}_{\tilde{I}_A})+2 j$, where $j$ is the winding number of the determinant of $\mathcal{U}(q)$. Hence, the winding number $W(\tilde{I}_A)$ represents a $\mathbb{Z}_2$-invariant, instead of a $\mathbb{Z}$-invariant. 
To simplify this expression, we use that Eq.~\eqref{eq:Isquared} implies
\begin{align*}
\frac{d}{dq}\left[\det{\mathcal{S}_{\tilde{I}_A}(q)}\det{\mathcal{S}_{\tilde{I}_A}(-q)}\right]=0.
\end{align*}
Hence, the integrand in Eq.~\eqref{eq:windingSIA} is even. As a result, we can write 
\begin{align*}
W(\mathcal{S}_{\tilde{I}_A})=\frac{i}{\pi}\left[\log{\det{\mathcal{S}_{\tilde{I}_A}(\pi)}}-\log{\det{\mathcal{S}_{\tilde{I}_A}(0)}}\right].
\end{align*}
This drastically simplifies the calculation of the $\mathbb{Z}_2$ invariant, since it frees us from the task of finding a smooth gauge over the full Brillouin zone. To define a $\mathbb{Z}_2$ invariant that takes values in the set $\{-1,1\}$, we introduce
\begin{align*}
\xi_{I_A}&:=e^{-i \pi W(\tilde{I}_A)}\\
&=\det{\mathcal{S}_{\tilde{I}_A}(\pi)}/\det{\mathcal{S}_{\tilde{I}_A}(0)}\nonumber\\
&=(-1)^{N_F}\det{\mathcal{S}_{\tilde{I}_B}(\pi)}/\det{\mathcal{S}_{\tilde{I}_B}(0)}\nonumber\\
&=:(-1)^{N_F}\xi_{I_B}
\end{align*}
In the third line we used that $\det{\mathcal{S}_{\tilde{I}_B}(\pi)}=\det{-\mathcal{S}_{\tilde{I}_A}(\pi)}$. 
We stress that these invariants do not depend in any way on the choice of unit cell or origin, and can therefore be considered as proper bulk invariants.

We can repeat the same analysis for mirror and rotation-symmetric insulators. The difference compared to inversion symmetry is that $\hat{M}_A^2=\hat{M}_B^2=(-1)^{2s}=\hat{C}_{2,A}^2=\hat{C}_{2,B}^2$. Here $\hat{M}_A$ ($\hat{C}_{2,A}$) and $\hat{M}_B$ ($\hat{C}_{2,B} $)are the mirror (rotation)-symmetry operators corresponding to mirror planes (rotation axes) $A$ and $B$, and $s$ is the total spin. However, this does not affect any of the above derivations.  Hence, for mirror-symmetric systems the winding number of the determinant of the sewing matrix $\mathcal{S}_{\tilde{M}_A}$ yields a $\mathbb{Z}_2$ classification. We characterize the parity of the winding number using the invariants $\xi_{M_A}$ and $\xi_{M_B}$, which are given by
\begin{align*}
\xi_{M_A}& :=e^{-i \pi W(\tilde{M}_A)}\\
&=\det{\mathcal{S}_{\tilde{M}_A}(\pi)}/\det{\mathcal{S}_{\tilde{M}_A}(0)}\nonumber\\
&=(-1)^{N_F}\det{\mathcal{S}_{\tilde{M}_B}(\pi)}/\det{\mathcal{S}_{\tilde{M}_B}(0)}\nonumber\\
&=:(-1)^{N_F}\xi_{M_B}.
\end{align*}
Similarly, we define the invariant $\xi_{C_{2,A}}$ corresponding to rotation symmetry. With this, we have seen that the sewing matrices play a key role within the topological classification of inversion-, rotation, or mirror-symmetric crystalline insulators. 

Let us now explore how the edge charge is related to these invariants. Considering inversion-symmetric systems and from the definition of the sewing matrices, it follows that 
\begin{align*}
|\Psi_m(-q)\rangle=\left[\mathcal{S}_{\tilde{I}_A}^*(q)\right]^{m,n}\tilde{I}_A(q)|\Psi_n(q)\rangle.
\end{align*}
Therefore, we obtain
\begin{align*}
&\Tr{\mathcal{A}(-q)}=\sum_n\langle\Psi_n(-q)|i\partial_{-q}|\Psi_n(-q)\rangle\nonumber\\
&=-\sum_{l,m,n}\langle\Psi_l(q)|\tilde{I}_A^\dagger(q)\left[\mathcal{S}_{\tilde{I}_A}(q)\right]^{n,l}i\partial_q\left[\mathcal{S}_{\tilde{I}_A}^*(q)\right]^{n,m}\tilde{I}_A(q)|\Psi_m(q)\rangle\nonumber\\
&=-\sum_n\langle\Psi_n(q)|i\partial_{q}|\Psi_n(q)\rangle-\Tr{\left[\mathcal{S}_{\tilde{I}_A}(q)i\partial_q\mathcal{S}^\dagger_{\tilde{I}_A}(q)\right]}\nonumber\\
&-\sum_n\langle\Psi_n(q)|\tilde{I}_A^\dagger(q)i\partial_q\tilde{I}_A(q)|\Psi_n(q)\rangle.
\end{align*}
Next, we note that 
\begin{align*}
\tilde{I}_A^\dagger(q)i\partial_q\tilde{I}_A(q)&=\begin{pmatrix}
0&0\\
0&\mathbb{I}_{n_\mathcal{B}\times n_\mathcal{B}}
\end{pmatrix}.
\end{align*}
As a result, we find
\begin{align}\label{eq:inversieBerry}
\Tr{\mathcal{A}(-q)}&=-\Tr{\mathcal{A}(q)}-\Tr{\left[\mathcal{S}_{\tilde{I}_A}(q)i\partial_q\mathcal{S}^\dagger_{\tilde{I}_A}(q)\right]}-\rho_\mathcal{B}(q)\nonumber\\
&=-\Tr{\mathcal{A}(q)}-i\frac{d}{dq}\log(\det \mathcal{S}^\dagger_{\tilde{I}_A}(q))-\rho_\mathcal{B}(q).
\end{align}
Here $\rho_\mathcal{B}(q)$ is the charge contained in partition $\mathcal{B}$. After integrating over all momenta, we finally obtain
\begin{align*}
\gamma&=i\log(\xi_{I_A})-\pi \rho_\mathcal{B}.
\end{align*}
Analogous expressions hold for rotation and mirror-symmetric insulators. Hence, in these systems we can express the edge charge as the sum of a topological and a non-topological part. Since the latter can be measured independently in the bulk, we conclude that the edge charge can indeed probe the topological $\mathbb{Z}_2$ invariant discussed above. Moreover, we stress that $\rho_\mathcal{B}$ depends continuously on external parameters, 
and therefore
any discontinuity in the edge charge can be only ascribed to a change in the band structure topology. 
Finally, we note that 
the edge charge can assume any value, except when the preferred unit cell 
is inversion-symmetric. Then, $\mathcal{B}=\varnothing$, which 
implies that the excess charge is quantized and given by $Q=i\log{(\xi_{X_A})}/2\pi$, with $X=I,M,\textrm{ or } C_2$.

Let us now elucidate these 
results by considering two examples. First, we study the binary chain. Here, we choose the unit cell with the red site at on-site energy $+m$.  Note that both the red and blue sites are inversion centers. 
We refer to the inversion center corresponding to the red sites as $A$. 
For this choice of unit cell, the Hamiltonian is given by
\begin{align*}
\tilde{H}(q)&=\begin{pmatrix}
m&t(1+e^{-iq})\\
t(1+e^{iq})&-m
\end{pmatrix}.
\end{align*}
Moreover, the inversion operator corresponding to the inversion center $A$ reads
\begin{align*}
\tilde{I}_A(q)&=\begin{pmatrix}
1&0\\
0&e^{-iq}
\end{pmatrix}.
\end{align*}
Hence, at half-filling we find that $\det{\mathcal{S}_{\tilde{I}_A}(0)}=1$, since the inversion operator is the identity matrix at $q=0$. For $q=\pi$, we have $\tilde{I}_A(0)=\sigma_3$. 
Therefore, we find $\det{\mathcal{S}_{\tilde{I}_A}(\pi)}=-\textrm{sign}(m/t)$. Combining these results, we have that $\xi_{I_A}=-\textrm{sign}(m/t)$, and therefore the edge charge $Q$ is given by
\begin{align*}
Q&=\begin{cases}1/2+\rho_B/2&\textrm{if sign}(m/t)>0\\
\rho_B/2&\textrm{if sign}(m/t)<0.
\end{cases}
\end{align*}
This result implies that 
 the jump in the edge charge encountered in Fig.~\ref{fig:fig3} follows from a change in the topology of  the band structure. 

Next, let us consider another well known toy-model that is inversion-symmetric: the Su-Schrieffer-Heeger(SSH) chain,\cite{SSH} depicted in Fig.~8. This chain consists of alternating solid and dashed bonds. These bonds are centers of inversion, which is in sharp contrast with the binary chain where the sites are inversion centers. We choose a unit cell with the dashed bond as its center. Since, the unit cell itself is inversion-symmetric we find that $\mathcal{B}=\varnothing$. Hence, for this  chain we expect a quantized edge charge. We refer to the solid bond as $A$, and the dashed bond joining the unit cells as bond $B$. The corresponding hopping parameters are denoted with $t_A$ and $t_B$. Using this notation, the Hamiltonian is given by
\begin{align*}
\tilde{H}(q)&=\begin{pmatrix}
0&t_A+t_B e^{-iq}\\
t_A+t_B e^{iq}&0
\end{pmatrix}.
\end{align*}
Under inversion, we find that the left and right sites are interchanged. Hence, the inversion operator is given by
\begin{align*}
\tilde{I}_A(q)&=\begin{pmatrix}
0&1\\
1&0
\end{pmatrix}.
\end{align*}
Note that the inversion operator is momentum independent, because all electrons belong to partition $\mathcal{A}$. At half-filling we find that $\xi_{I_A}=\textrm{sign}(1-t_B^2/t_A^2)$. Here, we have used that the lowest energy state, assuming $t_A$ to be a positive energy, at $q=0$ is given by $(1,-\textrm{sign}(1+t_B/t_A))^T/\sqrt{2}$ and at $q=\pi$ by $(1,-\textrm{sign}(1-t_B/t_A))^T/\sqrt{2}$. Hence, the edge charge is given by
\begin{align*}
Q&=\begin{cases}0&\textrm{if sign}(1-t_B^2/t_A^2)>0\\
1/2&\textrm{if sign}(1-t_B^2/t_A^2)<0.
\end{cases}
\end{align*}
We have numerically confirmed this result by computing the edge charge in a finite chain consisting of $100$ unit cells, with $t_A/t_B=2$($\xi_{I_A}=1$), and $t_A/t_B=1/2$($\xi_{I_A}=-1$). 
\begin{figure}[t]
\centering
\includegraphics[width=.475\textwidth]{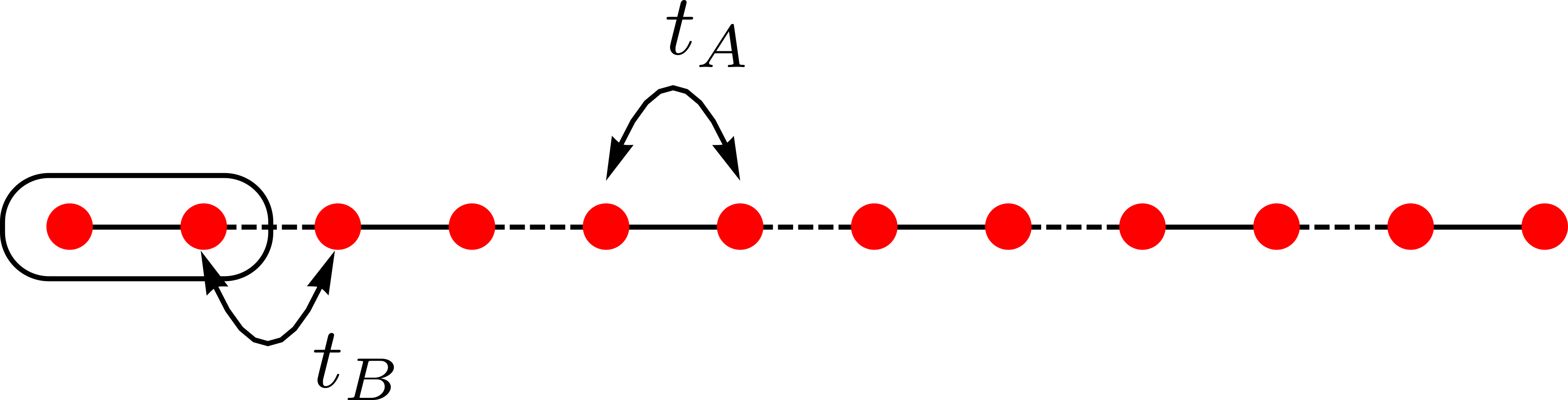}
\caption{Sketch of the SSH chain, the inter- (intra) unit-cell hopping is denoted with a dashed (solid) bond.}
\label{fig:fig8}
\end{figure}

Having established the generic relation between the excess charges and $\mathbb{Z}_2$ topological crystalline invariants, let us now consider the specific case of spin-one-half systems with time-reversal symmetry. 
First, we 
show that time-reversal symmetry implies that the topological invariants $\xi_{I_A}$, $\xi_{C_{2,A}}$, and $\xi_{M_A}$ introduced above are guaranteed to be trivial. 
When considering inversion-symmetric systems 
we can indeed write
\begin{align*}
\det{\mathcal{S}_{\tilde{I}_A}}(q)&=\prod_{j=1}^{N_F}\langle\Psi_j(q)|\tilde{I}_A(q)|\Psi_j(q)\rangle
=\prod_{j=1}^{N_F}\zeta_j(q),
\end{align*}
where $\zeta_j(q)$ denotes the eigenvalues of the sewing matrix and $|\Psi_j(0)\rangle$ the corresponding eigenstate. Then we find $\tilde{I}_A(0)\tilde{T}|\Psi_j(0)\rangle=\zeta^*(0)\tilde{T}|\Psi_j(0)\rangle$, and Kramer's theorem guarantees that these states are orthogonal. Hence, it follows that the $\det{S_{\tilde{I}_A}}(0)=1$. We can repeat the same argument for $q=\pi$. Therefore, we find $\xi_{I_A}=1=\xi_{I_B}$. 
This argument can be repeated for systems with two-fold rotation or mirror symmetry. 

Fortunately, this also offers new possibilities. Due to the $\pi$-periodicity of the determinant, 
for inversion-symmetric crystals
we might consider the winding number $W^{1/2}_{\tilde{I}_A}$ of $\det{\mathcal{S}_{\tilde{I}_A}}$ over half of the Brillouin zone, i.e. from $0$ to $\pi$
\begin{align}\label{eq:windingSIATRS}
W^{1/2}(\mathcal{S}_{\tilde{I}_A}):=\frac{i}{2\pi}\int_0^\pi\mathrm{d}q\frac{d}{dq}\log{(\det{\mathcal{S}_{\tilde{I}_A}(q)})}\in\mathbb{Z}.
\end{align}
When considering
an arbitrary gauge transformation, 
however, 
this winding number changes by an arbitrary integer. As such $W^{1/2}(\mathcal{S}_{\tilde{I}_A})$ has no meaning at all. However, if one imposes the time-reversal symmetric gauge, Eq.~\eqref{eq:TRSgauge}, then this winding number can only change by integer  multiples of $2$. To see this, let us suppose that we have found such a smooth time-reversal symmetric gauge. Then under a gauge transformation $|\Psi_m^\alpha(q)\rangle\rightarrow \mathcal{U}^{m,n}_{\alpha,\beta}(q)|\Psi_n^\beta(q)\rangle$. To respect the time-reversal symmetry constraint, one requires 
\begin{align}\label{eq:trsgt}
\mathcal{U}^{m,n}_{\alpha,\beta}(q)^*&=-\sum_{\gamma,\delta}\epsilon_{\alpha,\gamma}\mathcal{U}^{m,n}_{\gamma,\delta}(-q)\epsilon_{\delta,\beta}.
\end{align}
Where, $\epsilon_{\alpha,\beta}=-\epsilon_{\beta,\alpha}$, and $\epsilon_{I,II}=1$. This implies that $\det{\mathcal{U}^\dagger(q)}=\det{-\mathcal{U}(-q)}=\det{\mathcal{U}(-q)}$. Hence, under this gauge transformation we have $\det{S_{\tilde{I}_A}(q)}\rightarrow \det{S_{\tilde{I}_A}(q)}(\det{U(q)})^2$. Moreover, Eq.~\eqref{eq:trsgt} ensures that $\det{U(0)}=\det{U(\pi)}=1$. Combining these relations, we find $W^{1/2}_{\tilde{I}_A}\rightarrow W^{1/2}_{\tilde{I}_A}+2j$, with $j$ the winding number of the determinant of $\mathcal{U}$ over half of the Brillouin zone. \footnote{For completeness we note that one may drop the time-reversal symmetry constraint Eq.~\eqref{eq:TRSgauge}, by writing
\unexpanded{\begin{align*}
W^{1/2}(\mathcal{S}_{\tilde{I}_A})=\frac{i}{2\pi}\int_0^\pi\mathrm{d}q\frac{d}{dq}\log{(\det{\mathcal{S}_{\tilde{I}_A}(q)})}\nonumber\\
+\frac{i}{2\pi}\left[\int_0^\pi\mathrm{d}q\frac{d}{dq}\log{(\det{\mathcal{S}_{\tilde{T}}(q)})}-2\log\left(\frac{\Pf{\mathcal{S}_{\tilde{T}}(\pi)}}{\Pf{\mathcal{S}_{\tilde{T}}(0)}}\right)\right].
\end{align*}}
If a time-reversal symmetric gauge is employed the r.h.s. of the equation above reduces to Eq.~\eqref{eq:windingSIATRS}, and under an arbitrary gauge transformation it can only change by an integer multiple of $2$.} 
Analogously, to $\xi_{I_A}$ and $\xi_{I_B}$, we can then finally introduce the invariants $\chi_{I_A}$ and $\chi_{I_B}$:
\begin{align*}
\chi_{I_A}&:=e^{i\pi W^{1/2}_{\tilde{I}_A}}\\
&=(-1)^{N_F/2}e^{i\pi W^{1/2}_{\tilde{I}_B}}\\
&=:(-1)^{N_F/2}\chi_{I_B}
\end{align*}
These considerations also allow to define the $\mathbb{Z}_2$ topological crystalline invariants for 
rotation- and mirror-symmetric systems. However, there is a fundamental difference between these symmetries. Namely, for rotation and mirror-symmetric systems we find $(\hat{R}\hat{T})^2=(\hat{M}\hat{T})^2=1$, whereas for inversion-symmetric systems we find $(\hat{I}\hat{T})^2=-1$. As a consequence, 
in inversion-symmetric and time-reversal symmetric systems the bands are two-fold degenerate, 
whereas the band structure of mirror-symmetric and rotation-symmetric crystals   
generally 
exhibits degeneracies only at the time-reversal symmetric momenta $0$ and $\pi$. 
Kramer's theorem therefore ensures that in inversion-symmetric insulators the sewing matrix is block diagonal 
, {\it i.e.} $\langle\Psi^{II}_m(q)|\tilde{I}_A(-q)|\Psi^I(q)\rangle=0$, 
provided the time-reversal constraint Eq.~\eqref{eq:TRSgauge} is fulfilled. 
 Since the determinant of a block-diagonal matrix is the product of the determinants of the individual blocks, we 
  have $\det{S_{\tilde{I}_A}(q)}=\det{S^{I}_{\tilde{I}_A}(q)}\det{S^{II}_{\tilde{I}_A}(q)}$, which 
when using 
that $\langle\Psi^{II}_m(-q)|\tilde{I}_A(q)|\Psi^{II}_n(q)\rangle=\langle\Psi^I_n(-q)|\tilde{I}_A(q)|\Psi_m^I(q)\rangle$, yields
$\det{S_{\tilde{I}_A}(q)}=\det{S^{I}_{\tilde{I}_A}(q)}^2$.
Using Eq.~\eqref{eq:windingSIATRS}  we now find
\begin{align*}
W^{1/2}(\mathcal{S}_{\tilde{I}_A})&=\frac{i}{\pi}\int_0^\pi\mathrm{d}q\frac{d}{dq}\log{(\det{\mathcal{S}^I_{\tilde{I}_A}(q)})}\nonumber\\
&=\frac{i}{\pi}\left[\log{\det{\mathcal{S}^{I}_{\tilde{I}_A}(\pi)}}-\log{\det{\mathcal{S}^{I}_{\tilde{I}_A}(0)}}\right],
\end{align*}
which finally allows to express the $\mathbb{Z}_2$ topological invariant as
\begin{align*}
\chi_{I_A}&=\det{\mathcal{S}^{I}_{\tilde{I}_A}(\pi)}/\det{\mathcal{S}^{I}_{\tilde{I}_A}}(0).
\end{align*}
As a result, the crystalline topological invariant for inversion symmetric atomic chains can be computed using only the knowledge of the eigenstates at $q=0$ and $q=\pi$. 
This is different from rotation- and mirror-symmetric insulators where one has to find a smooth gauge in the full BZ. 

Let us now prove that the crystalline topological invariants 
$\chi_{I_A}$ and $\chi_{I_B}$ 
can be 
related to the partial Berry phase, 
which encodes the excess charge in time-reversal symmetric systems.  
Let us consider the time-reversal symmetric gauge, Eq.~\eqref{eq:TRSgauge}. This 
ensures that $\mathcal{A}(q)=\mathcal{A}(-q)$. 
If we combine this with Eq.~\eqref{eq:inversieBerry} we then find
\begin{align*}
\Tr{\mathcal{A}}(q)&=-\Tr{\mathcal{A}}(q)-i\frac{d}{dq}\log(\det \mathcal{S}^\dagger_{\tilde{I}_A}(q))-\rho_\mathcal{B}(q),
\end{align*}
which, when integrated from 
$0$ to $\pi$ 
yields the following relation between the partial Berry phase and the $\mathbb{Z}_2$ topological invariants:
\begin{align*}
\gamma^I=-i\log(\chi_{I_A})-\pi\rho_\mathcal{B}/2.
\end{align*}

\begin{figure}[t]
\centering
\includegraphics[width=.475\textwidth]{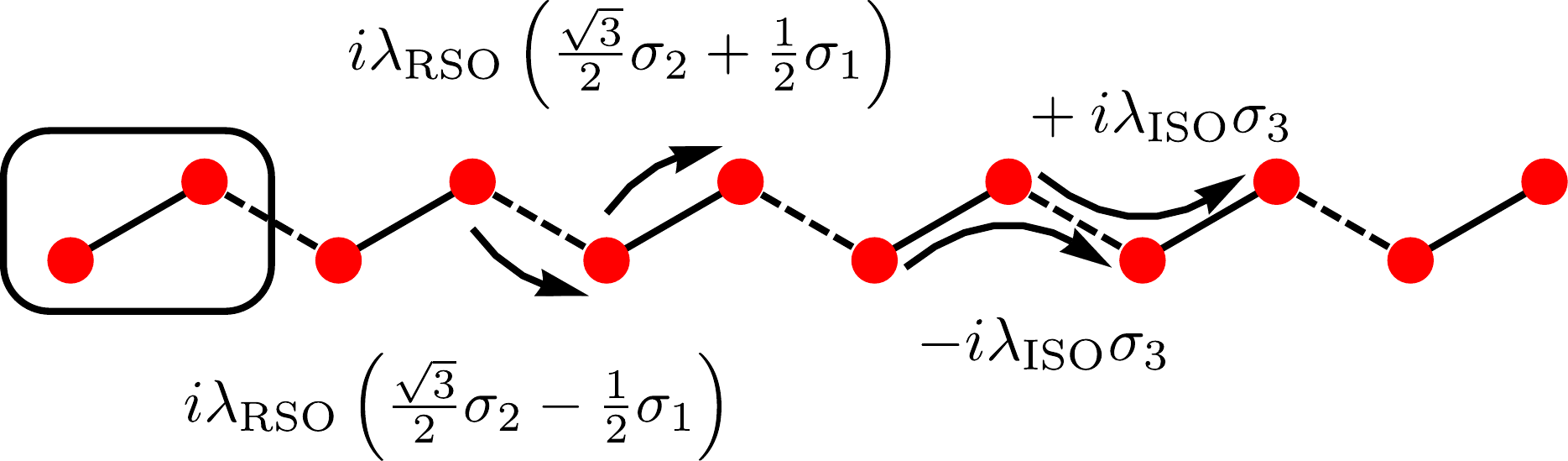}
\caption{Spinfull SSH chain, with $C_2$ symmetry. Spin-orbit coupling terms are schematically depicted with arrows.}
\label{fig:fig9}
\end{figure}
We now apply this result to a toy model, that can be seen as a 
spinful SSH atomic chain. In the absence of spin-orbit coupling the Hamiltonian is given by
\begin{align*}
\tilde{H}_0(q)&=\begin{pmatrix}
0&t_A+t_B e^{-iq}\\
t_A+t_B e^{iq}&0
\end{pmatrix}\otimes\sigma_0,
\end{align*}
where $\sigma_0$ is the identity operator acting in spin-space. 
Let us in addition
assume that the electrons are described by $p_z$-orbitals pointing out of the plane. 
We then 
find that the intrinsic spin-orbit coupling induces complex next-nearest neighbor hoppings, see Fig.~9. The corresponding Fourier transformed Hamiltonian term reads
\begin{align*}
\tilde{H}_\textrm{ISO}(q)&=\lambda_\textrm{ISO}\begin{pmatrix}
2\sin(q)&0\\
0&-2\sin(q)
\end{pmatrix}\otimes\sigma_3.
\end{align*}
Consequently, 
the full Hamiltonian reads $\tilde{H}(q)=\tilde{H}_0(q)+\tilde{H}_\textrm{ISO}(q)$. Since inversion-symmetry acts trivially in spin space, we find 
\begin{align*}
\tilde{I}_A(q)&=\begin{pmatrix}
0&1\\
1&0
\end{pmatrix}\otimes\sigma_0.
\end{align*}
Using that $\tilde{H}_\textrm{ISO}(0)=\tilde{H}_\textrm{ISO}(\pi)=0$, we find that the invariants for the spinful and spinless SSH chain are identical, i.e. $\chi_{I_A}=\textrm{sign}(1-t_B^2/t_A^2)$. Hence, the edge charge is given by $Q=i\log{(\textrm{sign}(1-t_B^2/t_A^2))}/\pi$. 
We have numerically verified this result by computing the edge charge of atomic chains of $100$ unit cells for the cases $t_A=2t_B=10\lambda_\textrm{ISO}$, and $t_B=2t_A=10\lambda_\textrm{ISO}$. 

We can also analyze the situation in which the
SSH chain lies on a substrate that breaks the out-of-plane reflection symmetry. This leaves us with a $2$-fold rotational symmetry around $A$ and $B$. 
The corresponding rotation operator is given by
\begin{align*}
\tilde{C}_{2,A}(q)&=\begin{pmatrix}
0&1\\
1&0
\end{pmatrix}\otimes i\sigma_3,
\end{align*}
where we used that a two-fold rotation around the $\hat{z}$ axis can be represented as $i\sigma_3$ in spin space. 
This rotation symmetry is preserved when we include a Rashba spin-orbit coupling due to the
broken mirror symmetry in the $\hat{z}$ direction. The Rashba spin-orbit coupling indeed yields  
nearest-neighbor hoppings accompanied by spin-flips (see Fig.~\ref{fig:fig9}) 
with an Hamiltonian term:
\begin{align*}
\tilde{H}_\textrm{RSO}(q)&=\lambda_\textrm{RSO,A}\begin{pmatrix}
0&1\\
-1&0
\end{pmatrix}\otimes \left(\frac{\sqrt{3}}{2}\sigma_2+\frac{1}{2}\sigma_1\right)\\
&+\lambda_\textrm{RSO,B}\begin{pmatrix}
0&e^{-iq}\\
-e^{iq}&0
\end{pmatrix}\otimes \left(-\frac{\sqrt{3}}{2}\sigma_2+\frac{1}{2}\sigma_1\right),
\end{align*}
such that the full Hamiltonian is given by $\tilde{H}_0(q)+\tilde{H}_\textrm{ISO}(q)+\tilde{H}_\textrm{RSO}(q)$. Now it is easily verified that $\tilde{C}_{2,A}(q)\tilde{H}(q)\tilde{C}_{2,A}(q)^{-1}=\tilde{H}(-q)$. Hence, we can then compute the $\mathbb{Z}_2$ invariant $\chi_{C_{2,A}}$. Here we find that for $t_A=2t_B=10\lambda_\textrm{ISO}=5\lambda_\textrm{RSO,A}=5\lambda_\textrm{RSO,B}$ a trivial invariant $\chi_{C_{2,A}}=1$, whereas for $t_B=2t_A=10\lambda_\textrm{ISO}=5\lambda_\textrm{RSO,A}=5\lambda_\textrm{RSO,B}$ we find $\chi_{C_{2,A}}=-1$. Indeed we find that in the former case the edge charge is trivial, whereas in the latter case a full electron is missing. 

Finally, let us consider the spinful binary chain discussed in Sec.~II.D. Inspection of Fig.~6(a) reveals that this system is mirror-symmetric. The corresponding symmetry operator is given by
\begin{align*}
\tilde{M}_A(q)&=\begin{pmatrix}1&0\\0&e^{-iq}\end{pmatrix}\otimes i\sigma_1.
\end{align*}
Next, we calculate the $\mathbb{Z}_2$-invariant $\chi_{M_A}$. At half-filling we find, using $(-1)m/t=10\lambda_\textrm{ISO}=10\lambda_\textrm{RSO}=1$, $\chi_{M_A}=(+1)-1$. Since, the bulk band gap only closes for $m=0$, we find that the edge charge is given by
\begin{align*}
Q&=\begin{cases}1+\rho_B/2&\textrm{if sign}(m/t)>0\\
\rho_B/2&\textrm{if sign}(m/t)<0.
\end{cases}
\end{align*}
Hence, the discontinuity in Fig.~\ref{fig:fig6}(c) can be attributed to a change in the crystalline topology.\\

\section{Conclusions}\label{sec:secconc}
To wrap up, we have shown that the excess charge in one-dimensional insulators, which do not carry a topological invariant according to the Altland-Zirnbauer classification, can be expressed in terms of the Berry phases of the bulk electronic Bloch waves. In presence of time-reversal symmetry, this relation can be conveniently expressed using the notion of the partial Berry phases. 
For atomic chains possessing spatial symmetries interchanging the chain ends, 
excess charges always contain a ``topological" contribution directly related to the $\mathbb{Z}_2$ invariants that can be associated with spatial-symmetric one-dimensional systems. 
Considering that one-dimensional topological crystalline insulating phases cannot be characterized by the presence of protected end modes -- these can be only stabilized by an additional non-spatial symmetry -- one can conclude that
the bulk-boundary correspondence can be only formulated in terms of  excess charges and that the latter can be then used to probe 
one-dimensional
crystalline topologies. 

\section{Acknowledgements}
We acknowledge the financial support of the Future and Emerging Technologies (FET) programme within
the Seventh Framework Programme for Research of the European Commission 
under FET-Open grant number: 618083 (CNTQC).  
C.O. acknowledges support from the Deutsche Forschungsgemeinschaft (Grant No. OR 404/1-1), and from a VIDI grant (Project 680-47-543) financed by the Netherlands Organization for Scientific Research (NWO). This work is part of the research programme of the Foundation for Fundamental Research on Matter (FOM), which is part of the Netherlands Organization for Scientific Research (NWO).

\appendix

\begin{widetext}

\section{Relation between the Zak phase and the Berry phase}
\label{sec:appendixa}
Here, we briefly show that the Berry phase as defined in Eq.~(2) corresponds to the inter-cellular part of the Zak phase $\gamma_\textrm{Zak}^\textrm{inter}$. The Zak phase $\gamma_\textrm{Zak}$ is expressed in terms of the cell-periodic part of the Bloch wave-function, which is given by
\begin{align}
u_{n,\alpha}(q)=e^{-i q r_\alpha}\Psi_{n,\alpha}(q).
\end{align} 
Here, $r_\alpha$ denotes the position of the $\alpha$-th orbital (spin, sub-lattice) within the unit-cell with respect to some reference point. With this, we find
\begin{align}
\gamma&=\sum_{n\leq N_F,\alpha}\int_{-\pi}^\pi\mathrm{d}qu^*_{n,\alpha}(q)e^{-i q r_\alpha}i\partial_q e^{i q r_\alpha}u_{n,\alpha}(q)=\sum_{n\leq N_F,\alpha}\int_{-\pi}^\pi\mathrm{d}qu^*_{n,\alpha}(q)i\partial_q u_{n,\alpha}(q)-\sum_{n\leq N_F,\alpha}\int_{-\pi}^\pi\mathrm{d}q\rho_{n,\alpha}(q)r_\alpha\nonumber\\
&=\gamma_\textrm{Zak}-\gamma^\textrm{intra}_\textrm{Zak}=:\gamma_\textrm{Zak}^\textrm{inter}
\end{align}

\section{Current operator}
\label{sec:appendixb}
Starting point is the Heisenberg equation of motion:
\begin{align}
\frac{d \hat{\rho}_i}{d t}&=i [\hat{H},\hat{\rho}_i]
\end{align}
Let us work out the commutator on the right-hand side. Keeping only terms in the Hamiltonian that contain $f^\dagger_i$ or $f_i$, we find
\begin{align}
[\hat{H},\hat{\rho}_i]&=\underbrace{\left[\sum_{j;\alpha,\beta} t_j^{\alpha,\beta}f^\dagger_{i,\alpha}f_{i+j,\beta},\hat{\rho}_i\right]}_{\textrm{B2.1}}+\underbrace{\left[\sum_{j\neq0;\alpha,\beta} t_{-j}^{\alpha,\beta}f^\dagger_{i-j,\alpha}f_{i,\beta},\hat{\rho}_i\right]}_{\textrm{B2.2}}.
\end{align}
Next, we work out both terms on the right-hand side by making use of the product rule. 
\begin{align}
\textrm{(B2.1)}&=\sum_{j;\alpha,\beta}t_j^{\alpha,\beta}\left\lbrace f^\dagger_{i,\alpha}\left[f_{i+j,\beta},\hat{\rho}_i\right]+\left[f^\dagger_{i,\alpha},\hat{\rho}_i\right]f_{i+j,\beta}\right\rbrace=\sum_{j;\alpha,\beta}t_j^{\alpha,\beta}\left\lbrace f^\dagger_{i,\alpha}\delta_{j,0}f_{i+j,\beta}-f^\dagger_{i,\alpha}f_{i+j,\beta}\right\rbrace=-\sum_{j\neq0;\alpha,\beta}t_j^{\alpha,\beta}f^\dagger_{i,\alpha}f_{i+j,\beta},
\end{align}
and
\begin{align}
\textrm{(B2.2)}&=\sum_{j\neq0;\alpha,\beta} t_{-j}^{\alpha,\beta}\left\lbrace f^\dagger_{i-j,\alpha}\left[f_{i,\beta}\hat{\rho}_i\right]+\left[f^\dagger_{i-j,\alpha},\hat{\rho}_i\right]f_{i,\beta}\right\rbrace=\sum_{j\neq0;\alpha,\beta} t_{-j}^{\alpha,\beta}f^\dagger_{i-j,\alpha}f_{i,\beta}.
\end{align}
Combining Eqs.~(B1), (B3), and (B4), we find
\begin{align}
\frac{d \hat{\rho}_i}{d t}&=i\sum_{j;\alpha,\beta}t_j^{\alpha,\beta}\left[f^\dagger_{i+j,\alpha}f_{i,\beta}-f^\dagger_{i,\alpha}f_{i+j,\beta}\right]=-\sum_j\hat{J}_{i\rightarrow i+j}.
\end{align}
Here we have defined the current operator as
\begin{align}
\hat{J}_{i\rightarrow i+j}&=i\sum_{\alpha,\beta}t_j^{\alpha,\beta}f^\dagger_{i,\alpha}f_{i+j,\beta}+h.c.
\end{align}
Next, we express this operator in terms of the Fourier transformed creation and annihilation operators
\begin{align}
\hat{J}_{i\rightarrow i+j}&=i\sum_{q\in BZ}\sum_{\alpha,\beta}t_j^{\alpha,\beta}e^{iqj}f^\dagger_{q,\alpha}f_{q,\beta}+h.c.+\ldots.
\end{align}
The dots correspond to term $f^\dagger_qf_{q'}$ with $q\neq q'$.  Now we turn to Eq.~\eqref{eq:currentop}, and write using the above result
\begin{align}
\hat{J}^\textrm{total}_{m\rightarrow m+1}&=\hat{J}_{m\rightarrow m+1}+\hat{J}_{m-1\rightarrow m+1}+\hat{J}_{m\rightarrow m+2}+\ldots=i\sum_{q\in BZ}\sum_{\alpha,\beta}\sum_{j\geq 0} jt_j^{\alpha,\beta}e^{iqj}f^\dagger_{q,\alpha}f_{q,\beta}+h.c.\nonumber\\
&=i\sum_{q\in BZ}\sum_{\alpha,\beta}\sum_{j} jt_j^{\alpha,\beta}e^{iqj}f^\dagger_{q,\alpha}f_{q,\beta}
\end{align}
The third equality follows from the Hermiticity of the Hamiltonian, i.e. $t_j^{\alpha,\beta}=(t_{-j}^{\beta,\alpha})^*$. We recognize that the summand is the derivative of the Hamiltonian $\tilde{H}(q)$, i.e.
\begin{align}
\hat{J}^\textrm{total}_{m\rightarrow m+1}&=\sum_{q\in BZ}f^\dagger_{q,\alpha}\nabla_q\tilde{H}^{\alpha,\beta}(q)f_{q,\beta}.
\end{align}

\end{widetext}


\begin{thebibliography}{10}
\bibitem{HasanKane}
\textrm{M.Z. Hasan and C.L. Kane}, {Rev. Mod. Phys.} {\bf 82}, 3045 (2010).
\bibitem{QiZhang}
\textrm{X.-L. Qi and S.-C. Zhang}, {Rev. Mod. Phys.} {\bf 83}, 1057 (2011).
\bibitem{FuKaneMele}
\textrm{L. Fu, C.L. Kane , and E.J. Mele}, {Phys. Rev. Lett.} {\bf 98}, 106803 (2007).
\bibitem{JoelMoore}
\textrm{J.E. Moore and L. Balents}, {Phys. Rev. B} {\bf 75}, 121306(R) (2007).
\bibitem{Fukane2}
\textrm{L. Fu and C. L. Kane}, {Phys. Rev. B} {\bf 76}, 045302 (2007).
\bibitem{SCZHANG1}
\textrm{H.J. Zhang, C.-X. Liu, X.-L. Qi, Z. Fang, and S.-C. Zhang}, {Nat. Phys.} {\bf 5}, 438 (2009).
\bibitem{SCZHANG2}
\textrm{C.-X. Liu, X.-L. Qi, H.J. Zhang, X. Dai, Z. Fang, and S.-C. Zhang}, {Phys. Rev. B} {\bf 82}, 045122 (2010).
\bibitem{VDBRINK}
\textrm{B. Rasche, A. Isaeva, M. Ruck, S. Borisenko, V. Zabolotnyy, B. B\"uchner, K. Koepernik, C. Ortix, M. Richter, and J. van den Brink}, {Nat. Mater.} {\bf 12}, 422 (2013).
\bibitem{LFU1}
\textrm{L. Fu}, {Phys. Rev. Lett.} {\bf 106}, 106802 (2011).
\bibitem{LFU2}
\textrm{T.H. Hsieh, H. Lin, J. Liu, W. Duan, A. Bansil, and L. Fu}, {Nat. Commun.} {\bf 3}, 982 (2012).
\bibitem{Tanaka}
\textrm{Y. Tanaka, Z. Ren, T. Sato, K. Nakayama, S. Souma, T. Takahashi, K. Segawa, Y. Ando}, {Nat. Phys.} {\bf 8}, 800 (2012).
\bibitem{VJ}
\textrm{R.-J. Slager, A. Mesaros, V. Juricic, and J. Zaanen}, {Nat. Phys.} {\bf 9}, 98 (2013).
\bibitem{AndoFu}
\textrm{Y. Ando and L. Fu}, {Ann. Rev. Condens. Matter Phys.} {\bf 6}, 361 (2015).
\bibitem{BalentsBurkov}
\textrm{A.A. Burkov and L. Balents}, {Phys. Rev. Lett.} {\bf 107}, 127205 (2011).
\bibitem{Savrasov}
\textrm{X. Wan, A.M. Turner, A. Vishwanath, and S.Y. Savrasov}, {Phys. Rev. B} {\bf 83}, 205101 (2011).
\bibitem{Hasan}
\textrm{S.-M. Huang, S.-Y. Xu, I.Belopolski, C.-C. Lee, G. Chang, B.K. Wang, N. Alidoust, G. Bian, M. Neupane, C. Zhang, S. Jia, A. Bansil, H. Lin, and M. Z. Hasan}, {Nat. Commun.} {\bf 6}, 7373 (2015).
\bibitem{Ding}
\textrm{B. Q. Lv, H. M. Weng, B. B. Fu, X. P. Wang, H. Miao, J. Ma, P. Richard, X. C. Huang, L. X. Zhao, G. F. Chen, Z. Fang, X. Dai, T. Qian, and H. Ding}, {Phys. Rev. X} {\bf 5}, 031013 (2015).
\bibitem{Hasan2}
\textrm{S.-Y. Xu, I. Belopolski, N. Alidoust, M. Neupane, G. Bian, C. Zhang, R. Sankar, G. Chang, Z. Yuan, C.-C. Lee, S.-M. Huang, H. Zheng, J. Ma, D.S. Sanchez, B.K.Wang, A. Bansil, F. Chou, P.P. Shibayev, H. Lin, S. Jia, and M. Z. Hasan}, {Science} {\bf 349}, 613 (2015).
\bibitem{Lau2}
\textrm{A. Lau, K. Koepernik, J. van den Brink, and C. Ortix}, {Phys. Rev. Lett.} {\bf 119}, 076801 (2017).
\bibitem{vonklitzing}
\textrm{K. von Klitzing, G. Dorda, and M. Pepper}, {Phys. Rev. Lett.} {\bf 45}, 494 (1980).
\bibitem{KaneMele}
\textrm{C. L. Kane and E. J. Mele}, {Phys. Rev. Lett.} {\bf 95}, 146802 (2005).
\bibitem{KaneMele2}
\textrm{C. L. Kane and E. J. Mele}, {Phys. Rev. Lett.} {\bf 95}, 226801 (2005).
\bibitem{BHZ}
\textrm{B.A. Bernevig, T.L. Hughes, and S.-C. Zhang}, {Science} {\bf 314}, 1757 (2006).
\bibitem{Molenkamp}
\textrm{M. K\"onig, S. Wiedmann, C. Br\"une, A. Roth, H. Buhmann, L.W. Molenkamp, X.-L. Qi, and S.-C. Zhang}, {Science} {\bf 318}, 766 (2007).
\bibitem{AZ}
\textrm{A. Altland and M.R. Zirnbauer}, {Phys. Rev. B} {\bf 55}, 1142 (1997).
\bibitem{Schnyder}
\textrm{A.P. Schnyder, S. Ryu, A. Furusaki, and A.W.W. Ludwig}, {Phys. Rev. B} {\bf 78}, 195125 (2008).
\bibitem{Kitaev}
\textrm{A. Kitaev}, {AIP Conf. Proc.} {\bf 1134}, 22 (2009).
\bibitem{Ryu}
\textrm{S. Ryu, A.P. Schnyder, A. Furusaki, and A.W.W. Ludwig}, {New. J. Phys.} {\bf 12}, 065010 (2010).
\bibitem{Friedel}
\textrm{J. Friedel}, {Philos. Mag.} {\bf 43}, 153 (1952).
\bibitem{Prodan}
\textrm{E. Prodan}, {Phys. Rev. B} {\bf 73}, 085108 (2006).
\bibitem{Loss}
\textrm{J.-H. Park, G. Yang, J. Klinovaja, P. Stano, D. Loss}, {Phys. Rev. B} {\bf 94}, 075416 (2016).
\bibitem{loss2}
\textrm{S. Gangadharaiah, L. Trifunovic, and D. Loss}, {Phys. Rev. Lett.} {\bf 108}, 136803 (2012).
\bibitem{loss3}
\textrm{P. Szumniak, J. Klinovaja, and D. Loss}, {Phys. Rev. B} {\bf 93}, 245308 (2016).
\bibitem{TKNN}
\textrm{D. J. Thouless, M. Kohmoto, M. P. Nightingale, and M. den Nijs}, {Phys. Rev. Lett.} {\bf 49}, 405 (1982).
\bibitem{Bardarson}
\textrm{J.W. Rhim, J. Behrends, J.H. Bardarson}, {Phys. Rev. B} {\bf 95}, 035421 (2017).
\bibitem{Vanderbilt}
\textrm{D. Vanderbilt and R. D. King-Smith}, {Phys. Rev. B} {\bf 48}, 4442 (1993).
\bibitem{kingsmith}
\textrm{R. D. King-Smith and D. Vanderbilt}, {Phys. Rev. B} {\bf 47}, 1651(R) (1993).
\bibitem{Zak}
\textrm{J. Zak}, {Phys. Rev. Lett.} {\bf 62}, 2747 (1989).
\bibitem{Bernevig}
\textrm{T.L. Hughes, E. Prodan, and B.A. Bernevig}, {Phys. Rev. B} {\bf 83}, 245132 (2011).
\bibitem{Chiu}
\textrm{C.-K. Chiu, H. Yao, and S. Ryu}, {Phys. Rev. B} {\bf 88}, 075142 (2013).
\bibitem{Shiozaki}
\textrm{K. Shiozaki and M. Sato}, {Phys. Rev. B} {\bf 90}, 165114 (2014).
\bibitem{Lau3}
\textrm{A. Lau, C. Ortix, J. van den Brink}, {Phys. Rev. Lett.} {\bf 115}, 216805 (2015).
\bibitem{Lau}
\textrm{A. Lau, J. van den Brink, and C. Ortix}, {Phys. Rev. B} {\bf 94}, 165164 (2016).
\bibitem{Fukane}
\textrm{L. Fu and C. L. Kane}, {Phys. Rev. B} {\bf 74}, 195312 (2006).
\bibitem{Thouless}
\textrm{D. J. Thouless}, {Phys. Rev. B} {\bf 27}, 6083 (1983).
\bibitem{Rigolin}
\textrm{G. Rigolin, G. Ortiz, and V. H. Ponce}, {Phys. Rev. A} {\bf 78}, 052508 (2008).
\bibitem{Kohn}
\textrm{W. Kohn}, {Phys. Rev.} {\bf 115}, 809 (1959).
\bibitem{Zaksym}
\textrm{J. Zak}, {Phys. Rev. B} {\bf 32}, 2218 (1985).
\bibitem{Niu}
\textrm{Q. Niu}, {Phys. Rev. B} {\bf 33}, 5368 (1986).
\bibitem{Fang}
\textrm{C. Fang, M. J. Gilbert, and B.A. Bernevig}, {Phys. Rev. B} {\bf 86} 115112 (2012).
\bibitem{Alexandrinata}
\textrm{A. Alexandradinata, X. Dai, B.A. Bernevig}, {Phys. Rev. B} {\bf 89}, 155114 (2014).
\bibitem{SSH}
\textrm{W. P. Su, J. R. Schrieffer, and A. J. Heeger}, {Phys. Rev. Lett.} {\bf 42}, 1698 (1979).
\end{thebibliography}
\end{document}